\documentclass[12pt]{iopart}
\usepackage{cite,amssymb,amsfonts,graphicx,bm,dcolumn,mathrsfs}
\newcommand{\bea}{\begin{eqnarray}}
\newcommand{\eea}{\end{eqnarray}}
\newcommand{\be}{\begin{equation}}
\newcommand{\ee}{\end{equation}}
\newcommand{\dd}{\mbox{d}}
\usepackage{epsfig}

\begin{document}

\title[Ising chains with competing interactions]{Ising chains with competing interactions
in presence of long--range couplings}

\author{Alessandro Campa$^1$, Giacomo Gori$^{2,3}$,
Vahan Hovhannisyan$^4$, Stefano Ruffo$^{5,6}$ and Andrea Trombettoni$^{3,5}$}
\address{$^1$ National Center for Radiation Protection and
Computational Physics, Istituto Superiore di Sanit\`{a},
Viale Regina Elena 299, 00161 Roma, Italy}
\address{$^2$ Dipartimento di Fisica e Astronomia G. Galilei,
Universit\'a degli studi di Padova, via Marzolo 8, 35131 Padova, Italy}
\address{$^3$ CNR-IOM DEMOCRITOS Simulation Center, Via Bonomea 265, I-34136
Trieste, Italy}
\address{$^4$ A. I. Alikhanyan National Science Laboratory, Alikhanian Br. 2, 0036 Yerevan, Armenia}
\address{$^5$ SISSA, via Bonomea 265, I-34136 Trieste, Italy \&
INFN, Sezione di Trieste, I-34151 Trieste, Italy}
\address{$^6$ Istituto dei Sistemi Complessi, CNR, via Madonna del Piano 10,
I-50019 Sesto Fiorentino, Italy}
\ead{\mailto{alessandro.campa@iss.it}, \mailto{gori@sissa.it}, \mailto{v.hovhannisyan@yerphi.am}, \mailto{ruffo@sissa.it} and \mailto{andreatr@sissa.it}}

\begin{abstract}
In this paper we study an Ising
spin chain with short--range competing interactions in presence of
long--range ferromagnetic interactions in the canonical ensemble. The simultaneous presence of the frustration
induced by the short--range couplings together with their competition
with the long--range term gives rise to a rich thermodynamic phase diagram. We compare
our results with the limit in which one of two local interactions is turned
off, which was previously studied in the literature. Eight regions of parameters
with qualitatively distinct properties are featured,
with different first-- and second--order phase transition lines and critical points.
\end{abstract}
\noindent{\it Keywords\/}: Phase transitions; long--range systems; competing interactions; phase diagram singularities.

\maketitle

\section{Introduction}
\label{intro}

The study of the effects of the presence of competing interactions
at different scales is of primary importance in many areas of physics.
In this respect, the competition of two
different forces, one attractive and one repulsive, is often considered. A vast amount of research
focused on either of two paradigmatic cases: {\it i)} the two forces act
on similar scales, giving possibly
rise to frustration \cite{Mezard87} since the particles
cannot minimize their energy without violating the constraints acting
on them;
{\it ii)} the two forces act on very different scales, one being much larger
than the other, which may result in the formation of patterns grown from instabilities
\cite{Chaikin95,Seul95}.

A paradigmatic context in which competing interactions are studied
is given by spin systems. There, one may have ferromagnetic and/or
antiferromagnetic couplings.
The ferromagnetic interactions favour alignment of spins, while the
antiferromagnetic ones tend to anti--align them. To exemplify the two
cases {\it i)} and {\it ii)} described above,
in the latter case {\it ii)} one may denote the large scale by $L$ (which
may be of the order of system size) and the small scale by $\ell$, with
$\ell \ll L$. When the long--range coupling is antiferromagnetic
and the short--range ferromagnetic, then the particles tend to align locally
and anti--align on larger distances, producing as an effect a rich variety
of ground states and in particular stripe patterns \cite{Seul95,Giuliani09}.
The correlation functions in the presence of competing long-- and
short--range
interactions and the resulting structure of multiple correlations and
modulation lengths have been deeply investigated (see {\it e.g.}
\cite{Chakrabarty11} and references therein).

Equally interesting is the case {\it i)}: denoting by $\ell_1$ and $\ell_2$
the length scales of the two competing couplings, frustration can emerge
when $\ell_1 \sim \ell_2$. In this Section we denote by $J_1$ and $J_2$
the strength of the coupling terms acting at scale $\ell_1$ and $\ell_2$,
respectively, and by $J$ the coefficient of the long--range coupling
acting at the scale $L$.
A typical, well studied example
is given by the so--called $J_1$--$J_2$ model, which has a nearest--neighbour (NN)
interaction $J_1$ and a next--nearest--neighbour (NNN) one $J_2$.
The properties of the $J_1$--$J_2$ model have been
thoroughly investigated both in classical and quantum spin models
\cite{Redner81,Mila91,Singh99,Capriotti04,Sirker06,Spenke12,Wang18}.
In one dimension the $J_1$--$J_2$
classical Ising model does not have a phase transition at finite temperature,
since one--dimensional short--range models never magnetize at $T>0$. However,
as a consequence of the competition between the two terms, this model does exhibit
infinite degeneracy of the ground state for a specific (negative)
value of the ratio
between $J_1$ and $J_2$ \cite{Redner81}.
In general, in order to have a magnetic transition
in one--dimensional classical models one needs long--range interactions
with power--law decay \cite{Dyson69,Thouless69,Sak73,Luijten02}.
The determination of the value of the
power--law exponent for which long--range interactions
become irrelevant with respect to short--range ones
has been recently at the center of intense investigations
\cite{Blanchard13,Brezin14,Angelini14,Defenu15,Defenu16,Gori17,Behan17,Morita17}.

In this paper, we aim at characterizing the effect of a double competition,
in which couplings at the different scales $\ell_1$, $\ell_2$ and $L$
are present,
with both $\ell_1 \sim \ell_2$ and $\ell_1,\ell_2 \ll L$. To illustrate
and work out the corresponding phases and thermodynamic phase diagram
in the simplest yet interesting setup,
we decided to consider a classical one--dimensional Ising
model with both NN and NNN couplings in presence of a long--range mean--field
ferromagnetic interaction. There are several reasons for such a choice.
First, the presence of all--to--all mean--field interactions considerably
simplifies the treatment, introducing
at the same time the effect of a long--range
interaction \cite{Campa14}. Second, when there is the long--range term in presence of a
single local NN interaction, the model has been solved exactly
in one dimension \cite{Nagle70,Kardar83}. In the limit
where both the long--range, mean--field--like, term and a single competing
short--range NN interaction are present ({\it i.e.}, the model with only $J_1$ and
$J$), it is known that the thermodynamic and dynamical behavior of the system
in both the canonical and microcanonical ensembles may be different
and one finds that
the two ensembles result in different phase diagrams \cite{Mukamel05}.
This inequivalence occurs in the region of parameters
where the long--range interaction is ferromagnetic ($J>0$)
and the short--range one is antiferromagnetic ($J_1<0$).
The model we consider in this paper
contains in itself the local
frustration induced by competing NN and NNN couplings, exhibiting at the same time
their competition with a mean--field ferromagnetic term,
and it has the advantage
of being exactly solvable in one dimension. We remind that in one dimension
there is no antiferromagnetic phase with a non vanishing staggered order
parameter at finite temperature. Moreover, models
of this kind should show more complex ground states
with respect to the standard $J_1$--$J_2$ model.

The plan of the paper is the following: In Section \ref{themodel} the model
studied in the rest of the paper is introduced and the solution in the
canonical ensemble of the case with only long--range and NN
couplings is reminded. In Section \ref{groundstate} the study of the
ground states of the model is presented, providing the basis for the
determination of the full phase diagram, which is then discussed in Section
\ref{phasediag}. In Section \ref{discussion} we provide a discussion of the main
features of the phase diagram, with the goal of presenting a synthetic qualitative
understanding of its richness. Our conclusions are presented in Section \ref{conclusions},
while more technical material is presented in the Appendix.

\section{The model}
\label{themodel}

We consider a one--dimensional lattice, where in each one of the $N$ sites there is an Ising spin variable $S_i$ with two
possible values, $+1$ and $-1$. The interactions between the spins are given by: an all--to--all mean--field ferromagnetic
coupling, a coupling between NN spins and a coupling between NNN spins. Following the notation of
Ref. \cite{Mukamel05} we decided to denote by $J$ the mean--field long--range coupling and by
$K_1$ the (ferro-- or antiferro--magnetic) NN coupling. The NNN coupling is then denoted as $K_2$, again possibly
positive or negative. Then, the Hamiltonian has the form:
\bea
H &=& -\frac{J}{2N} \sum_{i=1}^N \sum_{j=1}^N S_i S_j -\frac{K_1}{2}\sum_{i=1}^N S_i S_{i+1} -\frac{K_2}{2} \sum_{i=1}^N S_i S_{i+2}
\nonumber \\
&=& -\frac{J}{2N} \left(\sum_{i=1}^N S_i \right)^2 -\frac{K_1}{2}\sum_{i=1}^N S_i S_{i+1} -\frac{K_2}{2} \sum_{i=1}^N S_i S_{i+2} \, ,
\label{hamil}
\eea
where periodic boundary conditions are assumed.
When $J<0$ there is no order at finite temperature. Therefore we consider $J>0$
and, without loss of generality, we can take $J=1$, that formally amounts to measuring the
energy in units of $J$.

Depending on the sign of the other parameters, $K_1$ and $K_2$, we can have competing interactions. For example, while the mean--field
ferromagnetic interaction favours aligned spins, a negative value of $K_1$ would prefer NN with opposite alignments. Also when both $K_1$ and
$K_2$ are negative there is competition, since a negative $K_1$ prefers alternating spins, a configuration where NNN are aligned,
something that is not favoured by a negative $K_2$. We will see that these situations give rise to a very rich phase diagram.

It is useful to introduce the following order parameters:
\bea
\label{magn}
m &=& \frac{1}{N} \sum_{i=1}^N S_i \, ,\\
\label{g1near}
g_1 &=& \frac{1}{N} \sum_{i=1}^N S_i S_{i+1} \, ,\\
\label{g2next}
g_2 &=& \frac{1}{N} \sum_{i=1}^N S_i S_{i+2} \, ,
\eea
defining the average magnetization, the average NN correlation and the average NNN correlation. In terms of these
order parameters, the Hamiltonian can be written as
\be
\label{hamilorder}
H = - \frac{N}{2} \left( m^2 + K_1 g_1 + K_2 g_2 \right) \, .
\ee
We consider below the two limiting cases, $(K_1 \ne 0, K_2 = 0)$ and $(K_1 = 0, K_2 \ne 0)$.

\subsection{$K_2=0$}
\label{sub_K1}

When $K_2=0$, the model has been solved both in the canonical and microcanonical
ensembles \cite{Nagle70,Kardar83,Mukamel05}. Since in this paper we limit
ourself to the canonical ensemble, in this Section we provide a
brief reminder on the solution of the $K_2=0$ limit in the canonical ensemble.
Following \cite{Kardar83}, with the help of the Gaussian identity
(Hubbard--Stratonovich transformation)
\be
\label{HS}
\mathrm{exp} (ba^2)=\sqrt{\frac{b}{\pi}} \int^{+\infty}_{-\infty}
 dx \, \mathrm{exp} (-bx^2 +2abx) \, , \label{HST}
\ee
the partition function of the system,
$Z=\sum_{\{ S_i\}} e^{-\beta H}$, may be rewritten as
\be
\label{k20}
Z= \sqrt{\frac{ \beta N}{2\pi}}   \int_{-\infty}^{\infty} dx \, \sum_{\{ S_i\}}
\exp{\Big\{- \frac{N \beta x^2}{2} +\beta x \sum_{i=1}^N S_i+ \frac{\beta K_1}{2} \sum_{i=1}^NS_i S_{i+1}\Big\}}.
\ee
The sum on the spin configurations is readily performed giving
\be
\label{k21}
Z= \sqrt{\frac{ \beta N}{2\pi}}  \int_{-\infty}^{\infty} dx \, \exp{\Big\{- N \beta \Big[
\frac{x^2}{2} + f_0 \Big]
\Big\}} \, ,
\ee
where $f_0$ is the free energy density of the one--dimensional classical
NN Ising model with coupling $K_1/2$
in a magnetic field $x$ at temperature $T=1/\beta$
\cite{Parisi_book} (throughout the paper we use units in which $k_B=1$).
The free energy density of the model (\ref{hamil})
for $K_2=0$
is given by the saddle--point
method in the thermodynamic limit $N \to \infty$ by
$f=\min_x{\Big( \frac{\beta x^2}{2}+f_0\Big)}$.
One then finds \cite{Nagle70,Kardar83}
\be
\label{minf}
f=\min_x{\Bigg\{ \frac{\beta x^2}{2} -
\ln{\Big\{e^{\beta K_1/2} \, \cosh(\beta x) + \left[ e^{\beta K_1} \sinh^2(\beta x)+e^{-\beta K_1}\right]^{1/2}\Big\}}\Bigg\}} \, .
\ee
One finds that in the $(K_1,T)$ phase diagram there is a line of second--order phase
transitions with mean--field critical exponents; this line, $\beta_c(K_1)$, giving the critical value
of $\beta$ as a function of $K_1$, is defined implicitly by the equation
$\beta_c=e^{-\beta_c K_1}$ separating the disordered and
the ferromagnetic phases,
with vanishing and finite magnetization, respectively. This line exists
for $\beta_c<\beta_{TP}$ and it ends at a tricritical point given by $\beta_{TP} K_{1;TP}=-\frac{1}{2}\ln{3}$. Moreover,
as will be also discussed in Section \ref{phasediag}, further
decreasing the temperature one has a first--order phase transition
line reaching $T=0$ for $K_1=-1/2$.

\subsection{$K_1=0$}
\label{sub_K2}

The model with $K_1=0$, i.e.,
\be
H = -\frac{1}{2N} \left(\sum_{i=1}^N S_i \right)^2 -\frac{K_2}{2} \sum_{i=1}^N S_i S_{i+2} \, ,
\label{hamil_K10}
\ee
can be mapped in the thermodynamic limit to the model with $K_2=0$ and having
the value of $K_1$ equal to $K_2$.
To show it, let us consider the $K_1=0$ model (\ref{hamil_K10})
with even $N$ and free ends instead of periodic boundary conditions (this is irrelevant
in the thermodynamic limit).
Then, for $n=1,2,\dots,N/2$, we may pose
$S_{2n-1} \equiv \sigma_n$ and $S_{2n} \equiv \sigma_{\frac{N}{2}+n}$, so that
the above Hamiltonian (\ref{hamil_K10}) can then be written as
\be
H = - \frac{1}{2N} \left(\sum_{n=1}^{N} \sigma_n \right)^2 -\frac{K_2}{2} \sum_{n=1}^{N-1} \sigma _n \sigma_{n+1} + \frac{K_2}{2}
\sigma_{\frac{N}{2}} \sigma_{\frac{N}{2}+1} \, .
\ee
We then have the model (\ref{hamil}) with $K_2=0$ and $K_1=K_2$, apart
from one missing link, the one at the center of the lattice. Therefore in the
thermodynamical limit the free energy per site is the same, and so
the phase diagram of the model with $K_1=0$ in the $(K_2,T)$ plane is the same as the one
of the model with $K_2=0$ in the $(K_1,T)$ plane.

\section{The ground state}
\label{groundstate}

The evaluation of the ground state of the system is a very useful starting
point for the study of its thermodynamic phase diagram. In fact, the ground
state coincides with the equilibrium state at temperature $T=0$, and
its structure, as a function of the parameters
$K_1$ and $K_2$, provides
valuable hints for the determination of the equilibrium states at $T>0$.
The ground state as a function of $K_1$ and $K_2$ can be found by
considering the
possible ranges of variation of the order parameters. More precisely,
one can determine the state of minimum energy for a given value
of the magnetization $m$,
and then find for which value of $m$ one obtains the absolute minimum.
We can obviously restrict the study to $m\geq 0$, since the Hamiltonian
is even in $m$; more precisely, for each
configuration with $m>0$ there is another configuration with $m<0$
and with the same energy, obtained by reversing all spins.
This already implies that
a ground state with $m \neq 0$ is at least doubly degenerate.
For the evaluation of the possible ranges of the order parameters,
we will assume to study the system
in the thermodynamic limit $N\to \infty$. Then, for example, a configuration
with the first $N_1$ spins up and the following $N-N_1$ spins
down, gives $g_1=1$,
since the single pair of neighbouring spins with opposite alignments,
at the boundary between the two regions, gives a vanishing contribution to
$g_1$ in the
thermodynamic limit.

From Eq. (\ref{hamilorder}), the ground state is characterized by
finding the absolute minimum of the energy per particle
\be
\label{enerorder}
\epsilon \equiv H/N = - \frac{1}{2} \left( m^2 + K_1 g_1 + K_2 g_2 \right) \, ,
\ee
when the order parameters $m$, $g_1$ and $g_2$ vary over all their
possible values. Let us first determine what are these possible values.
Obviously the range of
variation of $m$ is between $-1$ and $1$. Then, we can evaluate
which is the range of $g_1$ for a given fixed value of $m$.
As underlined above, we can restrict to
$m \geq 0$; the possible ranges of $g_1$ and $g_2$ are the
same when $m \to -m$, since the values of $g_1$ and $g_2$ do not change by
reversing all spins. For a given
$m\geq 0$, the maximum value of $g_1$ is equal to $1$, occurring
when all up spins are grouped in a single cluster and
all down spins are grouped in another single
cluster. On the other hand, the minimum value is equal to $2m -1$,
occurring when all down spins are isolated
(this is possible since when $m\geq 0$ the number
of down spins is smaller than or equal to that of up spins).
By considering also negative values of $m$, one obtains that the
minimum value of $g_1$ is equal to
$2|m| -1$. We now consider the possible values of $g_2$ for
given values of $m\geq 0$ and $g_1$. The maximum value is equal to $1$,
which is achieved with
a configuration where the spins are divided in three clusters:
a cluster with a fraction $(1-g_1)/2$ of the spins with alternating
orientations, a cluster with
a fraction $(1+2m +g_1)/4$ of spins in the up state and a cluster
with a fraction $(1-2m +g_1)/4$ of spins in the down state
(we recall that $g_1 \ge 2|m| - 1$). For the evaluation of the minimum value of $g_2$
we can argue as follows. For a given value of $m \geq 0$
the minimum value of $g_2$, regardless of the
value of $g_1$, is $2m -1$, and it is obtained
when the spins are divided in two clusters, one with a fraction $(1-m)$ of
spins arranged with pairs of NN spins
alternatively up and down, and another cluster with a
fraction $m$ of spins in the up state
(again, by extending to negative values of $m$, one
finds that the minimum value of $g_2$ is $2|m| -1$). On the other hand,
to determine the minimum value of $g_2$ for a given value of $g_1$
and regardless of the value of $m$, we can write,
using that $S_i^2=1$, $\sum_i S_i S_{i+2} = \sum_i A_i A_{i+1}$,
where $A_i \equiv S_i S_{i+1}$. Then, the possible values
of $g_2$ for a given value of $g_1$ are the same as that of $g_1$ for
a given value of $m$. Therefore, for given $g_1$
the minimum value of $g_2$ is equal to
$2|g_1| -1$. Consequently, for given values of $m$ and $g_1$,
the minimum possible value of $g_2$ is $\max (2|m|-1,2|g_1|-1)$.

Let us summarize the allowed
ranges of the order parameters:
\be
\mkern-36mu \mkern-36mu
-1 \leq m \leq 1, \,\,\,\,\,\,\,\,\, 2|m|-1 \leq g_1 \leq 1, \,\,\,\,\,\,\,\,\,
\max(2|m|-1,2|g_1|-1) \leq g_2 \leq 1 \, .
\label{order_range}
\ee

\begin{figure}[h]
\begin{center}
\includegraphics[scale=0.7,trim= 0cm 3.5cm 0cm 0cm,clip]{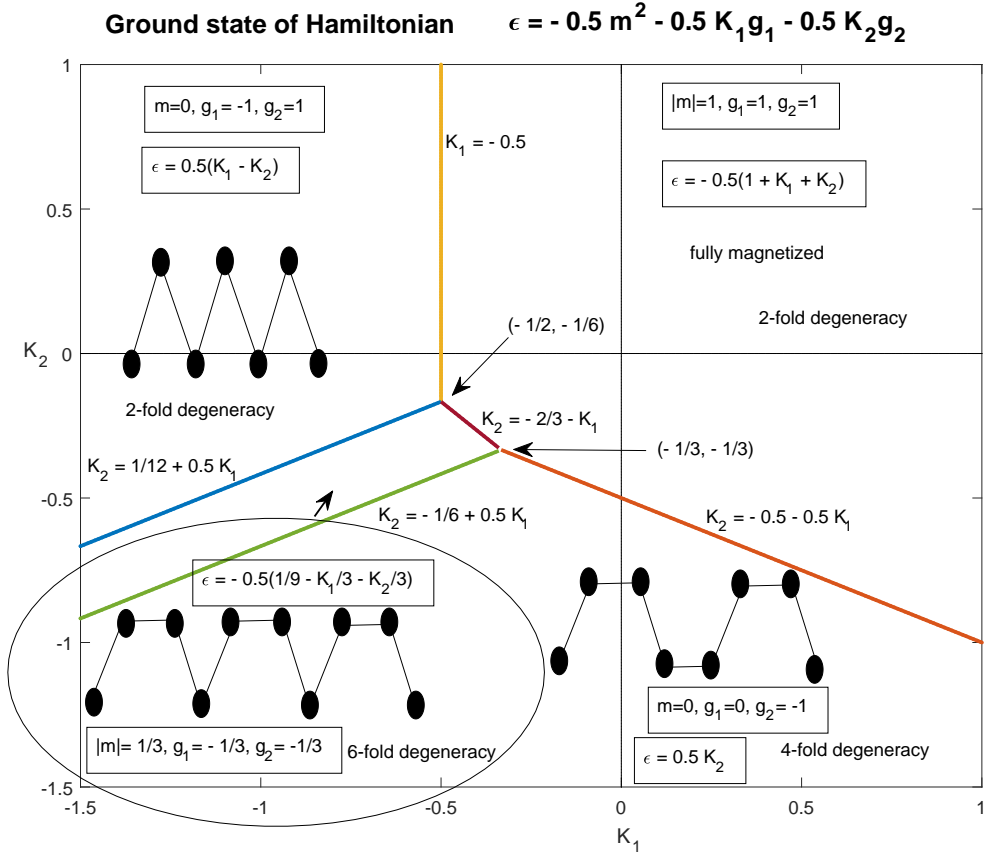}
\caption{Structure of the ground states in the $(K_1,K_2)$ plane. There are $4$ different structures, depending on the values of the parameters
$K_1$ and $K_2$, defined by the values of the order parameters $m$, $g_1$ and $g_2$. The colored lines divide the regions having ground states with
different structure; beside each line there is the equation defining it.
For each region the figure shows the values of the order parameters, the
corresponding expression of the energy per particle $\epsilon$, and the degeneracy of the ground state. Apart from the fully magnetized (ferromagnetic) state,
there are sketches showing the configuration of the partially magnetized
(ferrimagnetic) state and of the two paramagnetic ($m=0$) states; the segments
linking neighbouring spins are just for visual clarity.}
\label{fig_ground}
\end{center}
\end{figure}

The ground state of our system, for given values of $K_1$ and $K_2$,
is obtained when the order parameters $m$, $g_1$ and $g_2$ vary in the
allowed ranges.
In Appendix A we show some details of the evaluation, the results
of which are the following. The $(K_1,K_2)$ plane can be divided in
$4$ regions, where
the values of the order parameters $m$, $g_1$ and $g_2$
corresponding to the ground state are constant:
\begin{itemize}
\item In the region defined by the set of inequalities
\be
K_2 > - \frac{1}{2}K_1 -\frac{1}{2}, \,\,\,\,\,\,\,\,\,\,\,\,\,\,\,
K_2 > -K_1 -\frac{2}{3}, \,\,\,\,\,\,\,\,\,\,\,\,\,\,\,
K_1 > -\frac{1}{2} \, ,
\label{firstregion}
\ee
the equilibrium order parameters are
\be
\label{orderfirst}
|m|=1, \,\,\,\,\,\,\,\,\,\,\,\,\,\,\, g_1=1, \,\,\,\,\,\,\,\,\,\,\,\,\,\,\, g_2=1 \, ,
\ee
so that the energy per particle has the expression
\be
\label{enerfirst}
\epsilon = -\frac{1}{2} \left( 1 + K_1 + K_2 \right) \, .
\ee
This state is doubly degenerate, in correspondence of the
two possible spin orientations (see Figure \ref{fig_ground}).
\item In the region defined by the set of inequalities
\be
K_2 > \frac{1}{2}K_1 +\frac{1}{12}, \,\,\,\,\,\,\,\,\,\,\,\,\,\,\,
K_1 < -\frac{1}{2} \, ,
\label{secondregion}
\ee
the equilibrium order parameters are
\be
\label{ordersecond}
m=0, \,\,\,\,\,\,\,\,\,\,\,\,\,\,\, g_1=-1, \,\,\,\,\,\,\,\,\,\,\,\,\,\,\, g_2=1 \, ,
\ee
so that the energy per particle has the expression
\be
\label{enersecond}
\epsilon = \frac{1}{2} \left( K_1 - K_2 \right) \, .
\ee
This state has alternating up and down spins
(see Figure \ref{fig_ground}), and it is doubly degenerate.
The two degenerate states are obtained one
from the other by reversing all spins.
\item In the region defined by the set of inequalities
\be
K_2 < - \frac{1}{2}K_1 -\frac{1}{2}, \,\,\,\,\,\,\,\,\,\,\,\,\,\,\,
K_2 < \frac{1}{2} K_1 -\frac{1}{6} \, ,
\label{thirdregion}
\ee
the equilibrium order parameters are
\be
\label{orderthird}
m=0, \,\,\,\,\,\,\,\,\,\,\,\,\,\,\, g_1=0, \,\,\,\,\,\,\,\,\,\,\,\,\,\,\, g_2=-1 \, ,
\ee
so that the energy per particle has the expression
\be
\label{enerthird}
\epsilon = \frac{1}{2} K_2 \, .
\ee
This state has alternating pairs of up and down spins (see Figure \ref{fig_ground}), and it is $4$--fold degenerate. The four degenerate states
correspond to the four possible ways in which the spins labelled,
e.g., with $1$, $2$, $3$ and $4$, can be arranged.
\item In the region defined by the set of inequalities
\be
K_2 < \frac{1}{2}K_1 +\frac{1}{12}, \,\,\,\,\,\,\,\,\,\,\,\,\,\,\,
K_2 > \frac{1}{2}K_1 -\frac{1}{6}, \,\,\,\,\,\,\,\,\,\,\,\,\,\,\,
K_2 > -K_1 -\frac{2}{3} \, ,
\label{fourthregion}
\ee
the equilibrium order parameters are
\be
\label{orderfourth}
|m|=\frac{1}{3}, \,\,\,\,\,\,\,\,\,\,\,\,\,\,\, g_1=-\frac{1}{3}, \,\,\,\,\,\,\,\,\,\,\,\,\,\,\, g_2=-\frac{1}{3} \, ,
\ee
so that the energy per particle has the expression
\be
\label{enerfourth}
\epsilon = -\frac{1}{2} \left( \frac{1}{9} -\frac{1}{3} K_1 -\frac{1}{3} K_2 \right) \, .
\ee
This state has, for positive $m$, repeating triplets of spins with
two spins up and one down or two down and one up for negative $m$
(see Figure \ref{fig_ground}),
and it is $6$--fold degenerate; three degenerate states have positive $m$, and correspond to the three possible ways in which the spins labelled, e.g.,
with $1$, $2$ and $3$, can be arranged. The other three states have negative $m$ and are obtained by the first three by reversing all spins.
\end{itemize}
Figure \ref{fig_ground} summarizes the results for the ground states in
the plane $(K_1,K_2)$, also with the graphical sketches of the
equilibrium configurations.
The lines dividing the various regions in the $(K_1,K_2)$ plane are lines of
first--order phase transitions. A relevant fact is the presence, in a strip
of the $(K_1,K_2)$ plane, of a ground state
only partially magnetized (sometimes such states are called ferrimagnetic,
in contrast to the
fully magnetized ferromagnetic states). We note the presence of
two triple points, at the coordinates $(K_1,K_2)=(-1/2,-1/6)$ and
$(K_1,K_2)=(-1/3,-1/3)$, respectively. At the former there is equality of
the energies of the ferromagnetic state, the ferrimagnetic state,
and a paramagnetic ($m=0$) state. At the latter the equality
is of the energies of still the ferromagnetic and the ferrimagnetic state,
but the third state
is another paramagnetic state. Figure \ref{fig_ground} shows as well
the structure
of the two different paramagnetic states.
As discussed in the following Section,
the triple points are present also for a range of positive temperatures.

\section{Solution of the model and results for the phase
diagram} \label{phasediag}
\subsection{Transfer--matrix solution}

Let us now consider the solution of the model (\ref{hamil})
by means of the transfer
matrix method. The partition function can be written as
\be
Z = \sum_{\{S_{i}\}} e^{\frac{\beta}{2N} \left(\sum_{i=1}^N S_i \right)^2 +\frac{ \beta K_1}{2}\sum_{i=1}^N S_i S_{i+1} +\frac{ \beta K_2}{2} \sum_{i=1}^N S_i S_{i+2}} \, ,
\label{PF}
\ee
where $\beta=1/T$ and $\sum_{\{S_{i}\}}$ marks a summation over spin states.
With the help of the Gaussian identity (\ref{HS}),
the partition function of the system may be rewritten as
\be
Z = \sqrt{\frac{ \beta N}{2\pi}}
\int^{+\infty}_{-\infty} dx \, e^{ -\frac{\beta N}{2} x^2} \sum_{\{S_{i}\}} \prod_{i=0}^{N/2-1} T_{S_{2i+1},S_{2i+2}}^{S_{2i+3},S_{2i+4}} \,  ,
\label{PF1}
\ee
where
\bea
T_{S_1,S_2}^{S_3,S_4} &=& \exp \left\{ \frac{\beta x}{2} \left( S_1 + S_2 + S_3 + S_4 \right)
+ \frac{\beta K_1}{4} \left( S_1 S_2 + S_3 S_4 \right) \right. \nonumber \\ &+&\frac{\beta K_1}{2} S_2 S_3
+ \left. \frac{\beta K_2}{2} \left( S_1 S_3 + S_2 S_4 \right) \right\}.
\eea
We recall that we are assuming periodic boundary conditions, and besides we assume $N$ to be even; both assumptions,
useful for the computation, are physically irrelevant in the thermodynamic limit.
The partition function can be rewritten into the following form
\be
Z = \sqrt{\frac{ \beta N}{2\pi}}
\int^{+\infty}_{-\infty} dx \, e^{ -\frac{\beta N}{2} x^2} \, \mbox{Tr} \, T^{N/2} \,  .
\label{PF2}
\ee
Here $T$ is the transfer matrix which is formed by the
elements $T_{S_1,S_2}^{S_3,S_4}$ with on the rows the
elements $S_1$ and $S_2$ and on the columns the elements $S_3$ and $S_4$
taking values $\pm1$ (in this Section, and in some expressions of Section \ref{regA}, $T$ denotes the transfer matrix
rather than the temperature; this should not give rise to confusion). The transfer matrix $T$ reads
\be
T =  \left(
\begin{array}{cccc}
e^{ \beta \left( K_1 + K_2 + 2x \right)} &  e^{ \beta \left( \frac{K_1}{2} + x \right)} & e^{ \beta \left( -\frac{K_1}{2} + x \right)} &
e^{ -\beta K_2}  \\
e^{ \beta \left( -\frac{K_1}{2} + x \right)} & e^{ \beta \left( - K_1 + K_2 \right)} &  e^{ -\beta K_2} &
e^{ \beta \left( \frac{K_1}{2} - x \right)} \\
e^{ \beta \left( \frac{K_1}{2} + x \right)} &  e^{ -\beta K_2} & e^{ \beta \left( - K_1 + K_2 \right)} &
e^{ \beta \left( -\frac{K_1}{2} - x \right)} \\
e^{ -\beta K_2} & e^{ \beta \left( -\frac{K_1}{2} - x \right)} & e^{ \beta \left( \frac{K_1}{2} - x \right)} &
e^{ \beta \left( K_1 + K_2 - 2x \right)}
\end{array} \right).
\label{Talt}
\ee
$\mbox{Tr} T^{N/2}$ in the Eq.~(\ref{PF2})
means a trace of the matrix $T^{N/2}$ which can be
expressed through the eigenvalues of $T$ as
\be
 \mbox{Tr} T^{N/2} = \lambda_{1}^{N/2}+\lambda_{2}^{N/2}+\lambda_{3}^{N/2}+ \lambda_{4}^{N/2}.
\label{sumovereigen}
\ee
The eigenvalues of a nonsymmetric real matrix can be real or complex (appearing in complex conjugate pairs).
In our case we can rely on the Perron--Frobenius theorem \cite{Perron07,Frobenius12,Varga09}, that assures that
for a matrix with strictly positive elements, like our $T$, the eigenvalue with the largest modulus is real and
positive. Let us denote with $\lambda(x)$ (writing explicitly its dependence on $x$) this eigenvalue. From Eqs. (\ref{PF2})
and (\ref{sumovereigen}) it is clear that
in the thermodynamic limit the contribution of the other eigenvalues can be neglected.
The partition function of the system may be finally written as
\be
Z =  \sqrt{\frac{ \beta N}{2\pi}}
\int^{+\infty}_{-\infty} dx \, e^{ -N \beta \tilde{f} (\beta, x) } \,  ,
\label{PF3}
\ee
where
\be
\tilde{f} (\beta, x) = \frac{1}{2} x^2 - \frac{1}{2 \beta} \ln\lambda(x) \, .
\label{free_energy_ext}
\ee
The integral in Eq. (\ref{PF3}) can be performed by the saddle--point
method, and in the limit $N \rightarrow\infty$ one obtains
the following expression of the free energy per particle
\be
f(\beta) = \min_x{\Bigg\{ \frac{1}{2} x^2 - \frac{1}{2 \beta} \ln \lambda(x) \Bigg\}} \, .
 \label{free_en}
\ee
Although it is possible to write the analytical expression of $\lambda(x)$, since $T$ is a $4\times 4$ matrix, we have preferred to
obtain it numerically with the code that is also employed to find the minimum indicated in Eq. (\ref{free_en}).

\subsection{Phase Diagrams}

We are now in the position to discuss the phase diagram of the model.
We decided to show the phase diagram
in the two--dimensional $(K_1,T)$ plane for different values of $K_2$,
which is indeed
a convenient way to understand the rich structure of the phase diagram.
We preliminarily observe that the two limiting cases $K_2 \to 0$,
in which the long--range term competes with the NN
coupling $K_1$, and $K_1 \to 0$, with only long--range and the
NNN coupling $K_2$, have the same (two--dimensional) phase diagram.
Indeed, as discussed in Sections \ref{sub_K1} and \ref{sub_K2}, the model
with $K_1=0$ has the same free energy of the model with $K_2=0$ and $K_1=K_2$.
This, together with the need to have a visualization of the phase diagram
that overall depends on $K_1$, $K_2$ and $T$, suggests
to fix one of two couplings $K_1$, $K_2$ and vary the other at finite $T$.
Since the properties of the $K_2=0$ model are very well studied
\cite{Nagle70,Kardar83,Mukamel05}, we decided to fix $K_2$ and study the phase
diagram in the $(K_1,T)$ plane, analyzing how its structure depends on
the chosen fixed value of $K_2$.

A very rich structure of the phase diagram, coming from the interplay
between the competing interactions and the long--term coupling, emerges.
Depending on the values of $K_2$, we obtain
eight different phase diagrams qualitatively different between them.

In the following we study and show these eight regions separately,
and we postpone a qualitative discussion of the obtained findings
to Section \ref{discussion}.

\subsubsection{Region A: $K_2 > K_2^a \simeq -0.0885$\\}\label{regA}

In this region, where $K_2$ is larger than a value denoted by $K_2^a$, with
$K_2^a \simeq -0.0885$, the phase diagram in the $(K_1,T)$ plane
is qualitatively the same of the phase diagram for $K_2=0$, meaning
that the position
of the first--order and second--order transition lines depends explicitly on
$K_2$, but apart from that the phase diagram has the same form.

Let us first consider what happens at $T=0$. It can be seen
from Figure \ref{fig_ground} that the system in Region A
exhibits a first--order phase transition,
by increasing $K_1$, from a paramagnetic ($m=0$)
state to a ferromagnetic
one. By increasing $T$ the first--order phase transition
line changes to a second--order line.

The line of second--order transitions in the $(K_1,T)$
plane for given $K_2$ is in general
obtained by looking when the second derivative of the function (\ref{free_energy_ext}) in
$x=0$ vanishes. In addition one has to check that $x=0$ is
actually the absolute minimum of that function.

The function $\lambda(x)$ is given implicitly
by $\det |T(x) - \lambda {\rm I} |=0$, where
the solution of largest modulus has to be taken. It can be seen
that $T(x)$ is an even function of $x$,
and so is $\lambda(x)$. In fact, if after posing $x \to -x$
one first permutes the rows of $T$ according to $(1,2,3,4) \to (4,3,2,1)$
and then
makes the same permutation in the columns, one has again $T(x)$.
Since the above permutations do not alter the determinant, this shows that
$\det |T(-x) - \lambda {\rm I} | = \det |T(x) - \lambda {\rm I} |$ and
$\lambda(-x) = \lambda(x)$, as one could have guessed on physical basis.
Then, all odd derivatives of $\det |T(x) - \lambda {\rm I} |$ in $x=0$, and
consequently of $\lambda(x)$, vanish.

To proceed, we observe that if $F(x,y)=0$ defines implicitly the
function $y(x)$, then, using the usual notation for partial
derivatives, we have
\bea
\frac{\dd y}{\dd x} \equiv y_x &=& - \frac{F_x}{F_y} \, , \\
\frac{\dd^2 y}{\dd x^2} \equiv y_{xx} &=& -\frac{F_{xx} + 2F_{xy} y_x + F_{yy} y_{x}^2}{F_y} \, .
\eea
In our case $F= \det |T(x) - \lambda {\rm I} |$,
with $\lambda$ playing the role of $y$. We have seen that
$F_x=F_{xy}=y_x=0$, so that
\be
\frac{\dd^2 \lambda}{\dd x^2} = -\frac{F_{xx}}{F_{\lambda}} \, .
\ee
Let us denote with $T^{(p)}$ the determinant of $T$ after the substitution
of the $p$-th column with the derivative of its elements
with respect to $x$. For example, in $T^{(1)}$ the first column of
$T$ is substituted by
$\left[ 2\beta e^{ \beta \left( K_1 + K_2 + 2x \right)}, \beta e^{ \beta \left( -\frac{K_1}{2} + x \right)},
\beta e^{ \beta \left( \frac{K_1}{2} + x \right)}, 0 \right]^T$.
In the same way, we denote with $T^{(p,q)}$ the determinant of $T$
after the substitution: ($i$) if $p\ne q$, of the $p$-th column with the derivative of its elements with respect to $x$ and of
the $q$-th column with the derivative of its elements with respect to $x$; ($ii$) if $p=q$, of the $p$-th column with the second derivative
of its elements with respect to $x$. Then, we have:
\be
F_{xx} = \sum_{p=1}^4 \sum_{q=1}^4 T^{(p,q)} \, .
\ee
Furthermore, we denote with $T_{(p)}$ the determinant of $T$ after the substitution of the $p$-th column with a column with $-1$
in the $p$-th row and $0$ in the other three rows. Then
\be
F_{\lambda} = \sum_{p=1}^4 T_{(p)} \, .
\ee
So finally
\be
\left. \frac{\dd^2 \lambda}{\dd x^2} \right|_{x=0} \equiv \lambda^{(2)} = \left . - \left\{ \left[ \sum_{p=1}^4 \sum_{q=1}^4 T^{(p,q)}\right]
\left[ \sum_{p=1}^4 T_{(p)}\right]^{-1} \right\} \right|_{x=0} \, .
\ee
For $x\to 0$, we can take $\lambda(x) = \lambda(0) + \frac{1}{2}\lambda^{(2)}x^2$, and then $\ln \lambda(x) = \ln \lambda(0)
+ \frac{1}{2\lambda(0)}\lambda^{(2)}x^2$. Then,
the second derivative (with respect to $x$)
in $x=0$ of the expression in curly brackets
in Eq. (\ref{free_en}) is given by
$1 - \lambda^{(2)}/[2 \beta \lambda(0)]$.
By numerically finding where the latter expression vanishes
in the $(K_1,T)$ plan, one obtains the points potentially belonging to the
line of second--order transitions. The points among them that actually belong to the line are those that, in addition, are an absolute minimum of
the expression in the brackets in Eq. (\ref{free_en}). Therefore,
from the operative point of view, to determine the second--order line,
the minimum search implied in Eq. (\ref{free_en}) has to be
performed only for the $(\beta,K_1)$ values where $1 - \lambda^{(2)}/[2 \beta \lambda(0)]$ vanishes.
Finally, to complete the determination of the
phase diagram, one has to determine the locus of points in which the
magnetization value has a discontinuity, therefore obtaining the first--order
phase transition line.

\begin{figure}[!h]
\begin{center}
\begin{tabular}{ccc}
\includegraphics[width=8cm]{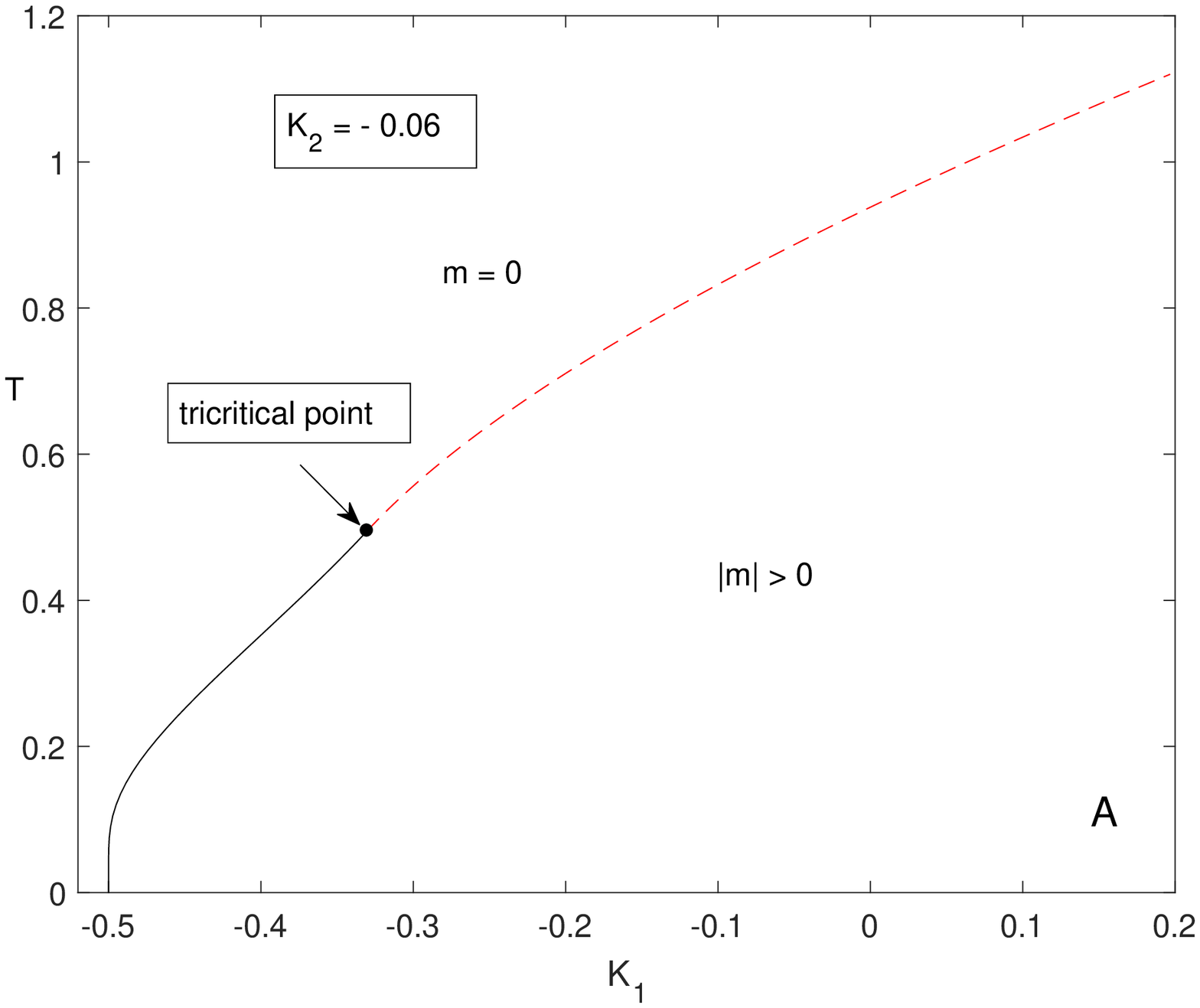} &
\includegraphics[width=8cm]{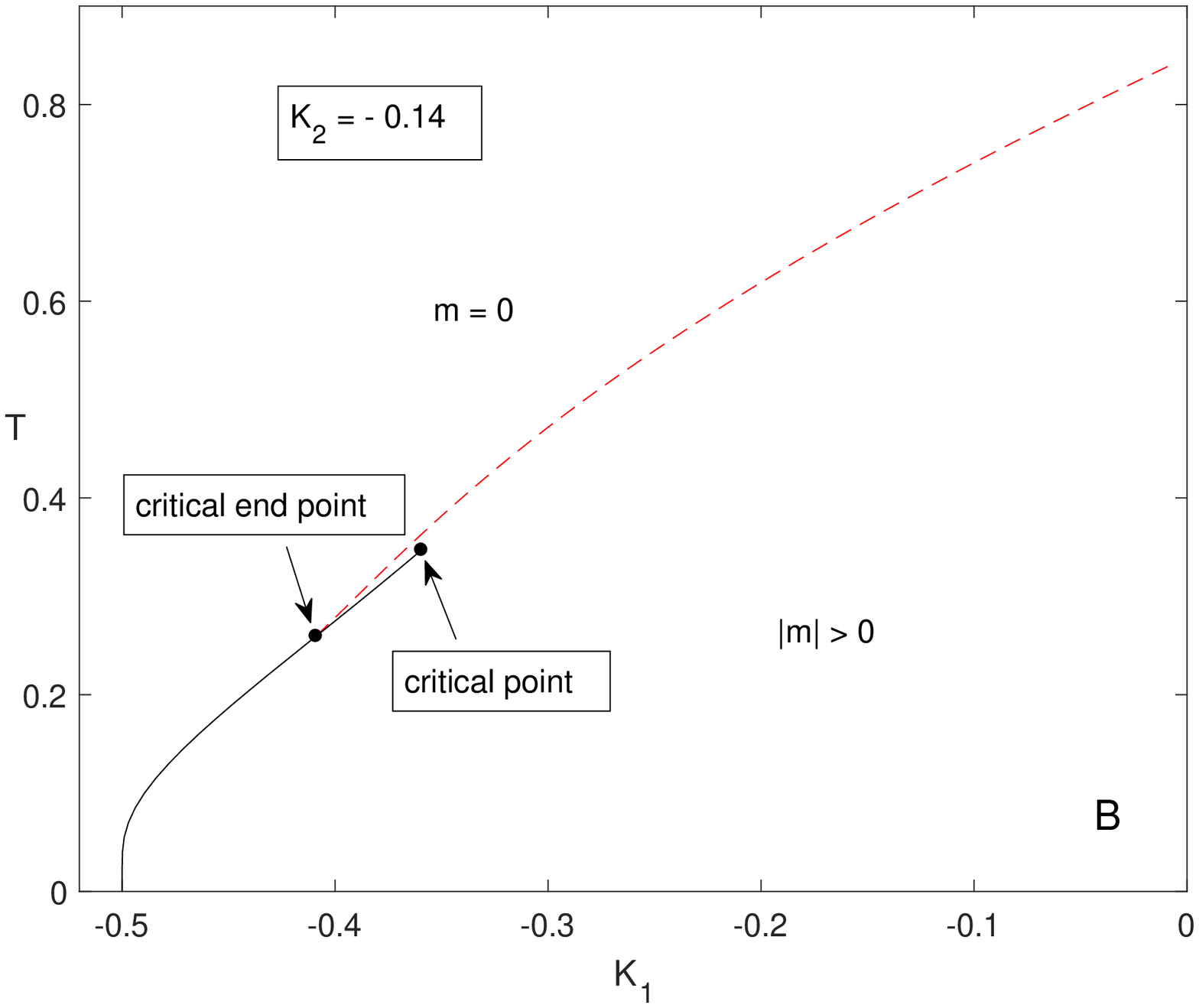}
\end{tabular}
\caption {(left)$(K_1, T)$ phase diagram of the model corresponding
to the value $K_2=-0.06$ in the region $A$. In this, and in all the other
plots of phase diagrams, solid and dashed lines
are the first-- and second--order phase transition lines, respectively.
(right) Phase diagram for $K_2=-0.14$ in the region $B$.} \label{Phase1}
\end{center}
\end{figure}

The corresponding results are plotted in Figure \ref{Phase1} (left) for a value
of $K_2$ in the region A ($K_2=-0.06$). It is seen that when the temperature decreases
the second--order line ends at a tricritical point, and that from there
a first--order line originates, going up to $T=0$. For the value of $K_2$ chosen
in Figure \ref{Phase1} (left) the
coordinates of the tricritical point are
$(K_{1} \backsimeq -0.330, T \backsimeq 0.495$).
It has to be observed that the phase diagram plotted in Figure \ref{Phase1} (left) is qualitatively the same as the canonical phase diagram
discussed in \cite{Kardar83,Mukamel05} for $K_2=0$.\\

\subsubsection{Region B:
$ -0.1542 \simeq K_2^b <K_2 < K_2^a \simeq -0.0885$\\}\label{regB}

When the NNN coupling $K_2$ is further decreased
below the critical value $K_2^a$, then below the second--order transition
line a first--order line emerges. This implies that the second--order
line terminates at a critical end point, while the first--order line
starting at $T=0$ ends at a critical point. The section of the first--order line between the critical end point
and the critical point separates two different magnetized phases (see right panel of Figure \ref{Phase1}). As usual in the presence
of a critical point ending a first--order transition line, the two phases could be connected by a path in the $(K_1,T)$ phase diagram that
does not intersect the transition line, so that moving on the path one does not encounter discontinuities in the magnetization.
This peculiar phase diagram,
obtained following the procedure described in Section \ref{regA},
occurs when $K_2$ is between a value $K_2^b \simeq -0.1542$ and $K_2^a$. A typical phase diagram for a value
of $K_2$ in the region B ($K_2=-0.14$) is reported in Figure
\ref{Phase1} (right), where
the coordinates of the critical end point and of the critical points
are respectively $(K_{1} \backsimeq -0.409, T  \backsimeq 0.259)$ and
$(K_{1} \backsimeq -0.359, T \backsimeq 0.347$).

\subsubsection{Region C: $ -1/6 <K_2 < K_2^b \simeq -0.1542$\\}\label{regC}

Below $K_2^b$, in the region C defined by $-1/6 <K_2 < K_2^b \simeq -0.1542$,
the first--order line bifurcates in a triple point. One of the two
first--order line ends at a critical point, as in region B, but the other
terminates at a new tricritical point where it meets
the second--order line. The phase diagram for the value $K_2=-0.16$ in the
region C is reported in Figure \ref{Phase2} (left), where
the coordinates of the tricritical point, of the triple point and of the critical
point are respectively
($K_{1} \backsimeq -0.499, T  \backsimeq 0.095 $),
($K_{1} \backsimeq -0.493, T  \backsimeq 0.076$) and
($K_{1} \backsimeq -0.355, T  \backsimeq 0.329$).

\begin{figure}[!h]
\begin{center}
\begin{tabular}{ccc}
\includegraphics[width=8cm]{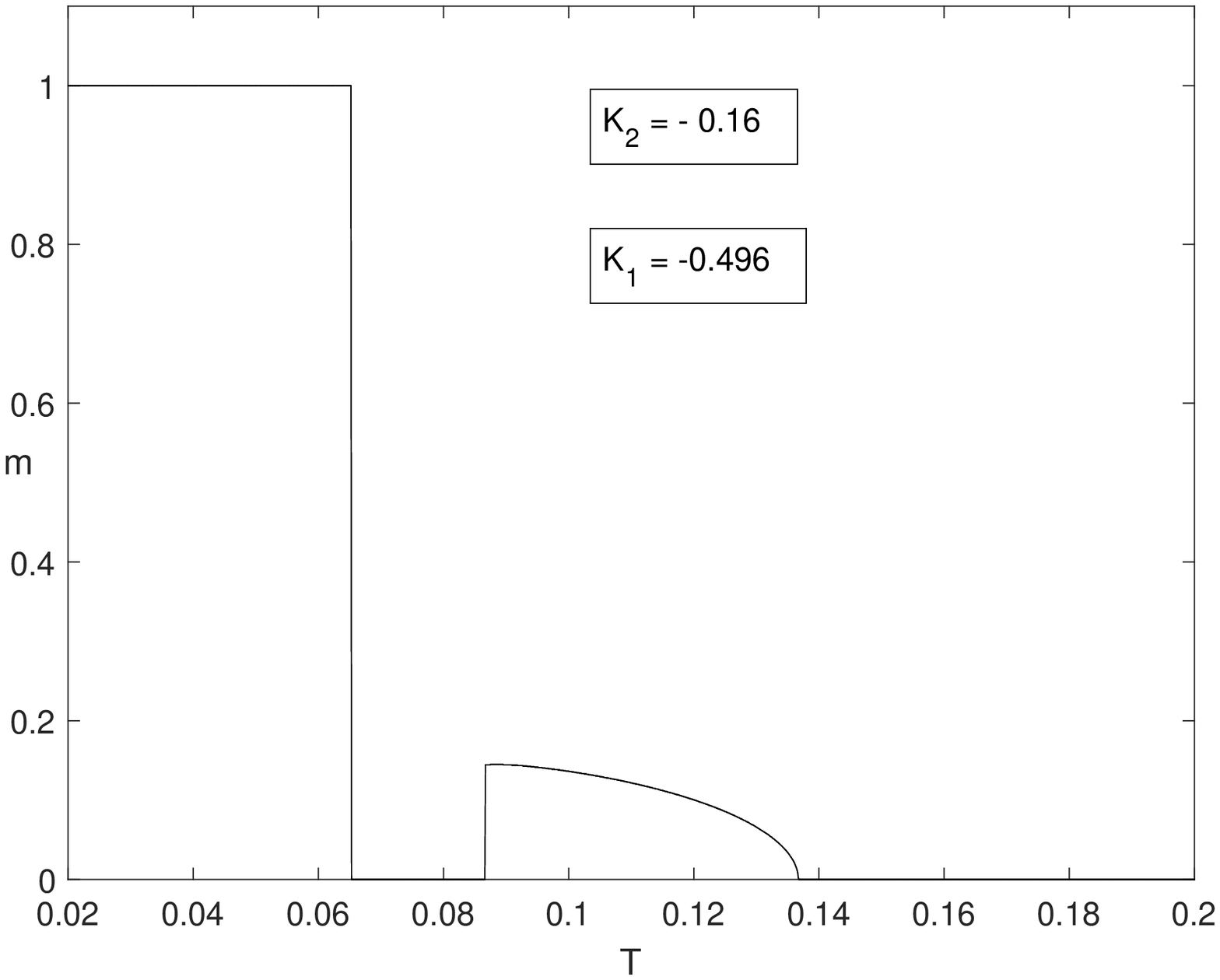} &
\includegraphics[width=8cm]{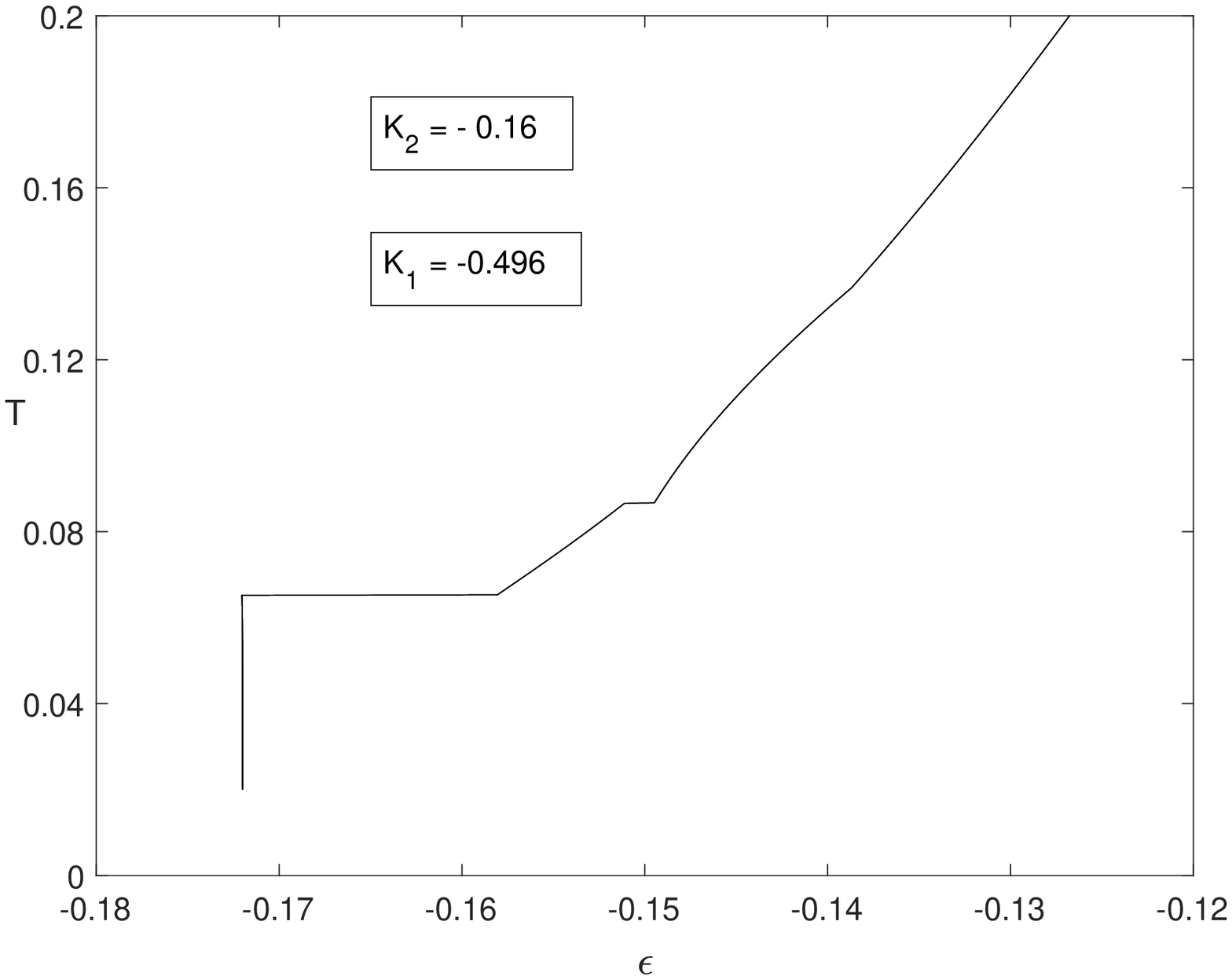}
\end{tabular}
\caption {(left) Magnetization {\it vs} temperature for $K_1=-0.496$ and
$K_2=-0.16$ in region C. (right) Temperature {\it vs} energy per particle (caloric curve)
for the same value of the parameters. The two flat regions are the Maxwell constructions
in correspondence of the two first--order phase transitions; the discontinuity of the derivative
in correspondence of the second--order phase transition at $\epsilon \backsimeq -0.139$ and
$T \backsimeq 0.137$ is hardly visible.} \label{MagnPhase2}
\end{center}
\end{figure}

To better illustrate the behaviour of thermodynamic quantities
in region C, we plot in Figure \ref{MagnPhase2} the magnetization
as a function of temperature (left) and
the temperature as a function of the energy per particle (right)
for the same value of $K_2$ chosen in Figure \ref{Phase2}, $K_2=-0.16$.
$K_1$ is chosen so that, fixing both $K_2$ and $K_1$ and increasing
$T$ one passes through the first--order line connecting the triple
point and the tricritical point. One clearly see in Figure \ref{MagnPhase2}
(left) that with $K_1=-0.496$, increasing $T$ the magnetization drops down to $0$, it is again
different from zero with a jump, and then go smoothly to zero in correspondence
of the second--order phase transition. In Figure \ref{MagnPhase2} (right) we have plotted the
so--called caloric curve (temperature $T$ {\it vs} the energy per particle $\epsilon$),
in which the two flat regions show the energy jumps in correspondence of the
two first--order phase transitions (Maxwell construction).

\subsubsection{Region D: $ -0.2672 \simeq K_2^c <K_2 < -1/6$\\}\label{regD}

The behaviour of region C changes at $K_2=-1/6$, as one can understand
by looking at the $T=0$ ground state diagram of Figure \ref{fig_ground}. For
$K_2>-1/6$, at low temperature for $T \to 0$ there are only two phases:
one having magnetization $m=0$ and the other $|m|>0$, and by fixing a
small enough $T$ and $K_2>-1/6$
one can pass from one region to the other by a first--order transition.
The situation changes for $K_2<-1/6$, where {\it two} first--order lines
start from $T=0$. In other words, the triple point of region C
occurs at a decreasing temperature when $K_2$ decreases,
and when it reaches $T=0$ at $K_2 = -1/6$, then it gives rise
to two first--order lines by further decreasing $K_2$. It is very interesting to notice that while one of two
lines turns right increasing the temperature,
{\it i.e.} $dT_c^{I;a}/dK_1>0$, the new first--order line
turns left, {\it i.e.} $dT_c^{I;b}/dK_1<0$ (where we denote by $T_c^{I;a}$
and $T_c^{I;b}$ the two first--order lines, where the $''a''$--one is the one
for larger values of $K_1$). Furthermore, the second--order line reaches
the $''b''$ first--order line in a tricritical point, so that in region D the
phase diagram features a critical point and a tricritical point.

\begin{figure}[!h]
\begin{center}
\begin{tabular}{ccc}
\includegraphics[width=8cm]{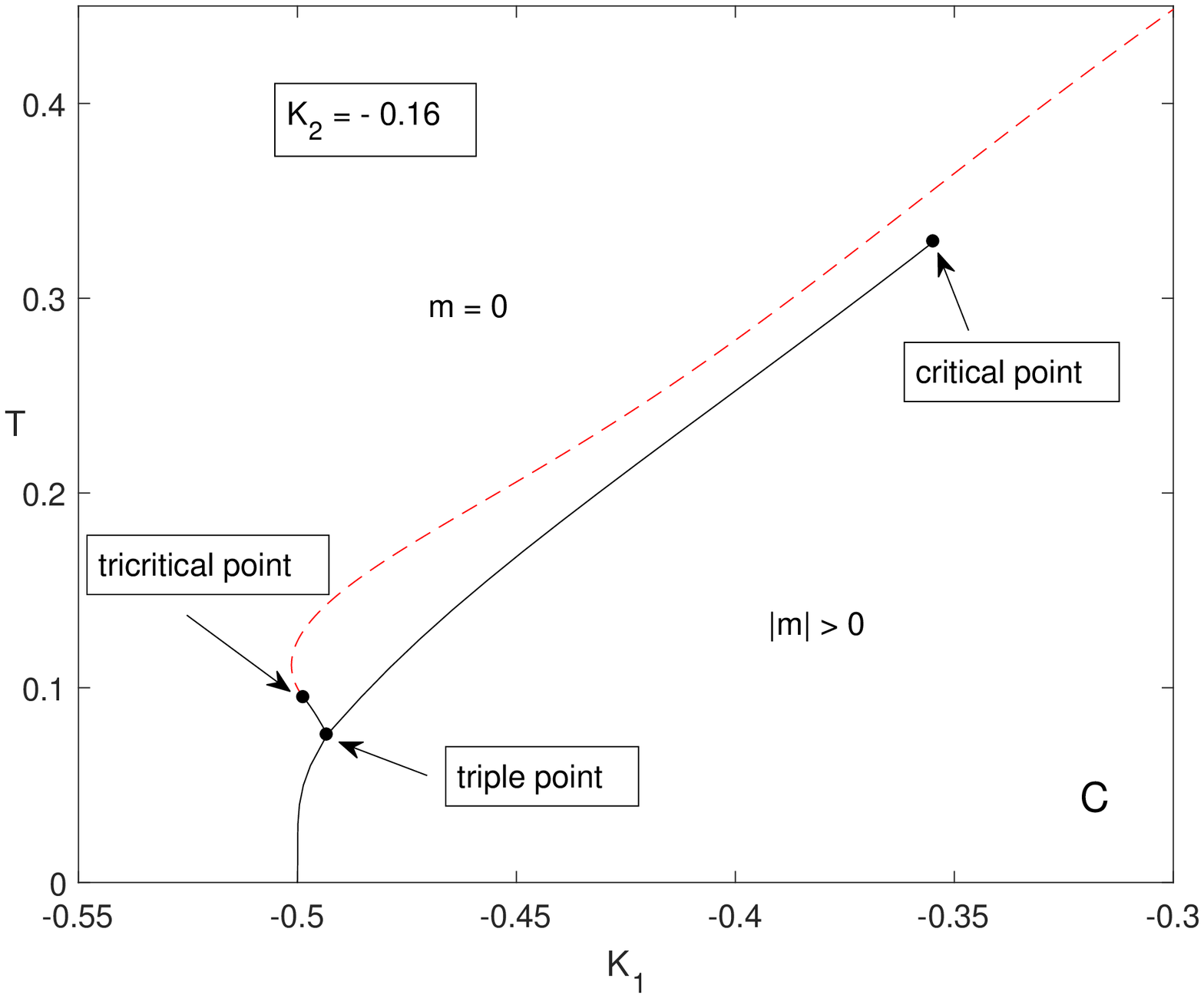} &
\includegraphics[width=8cm]{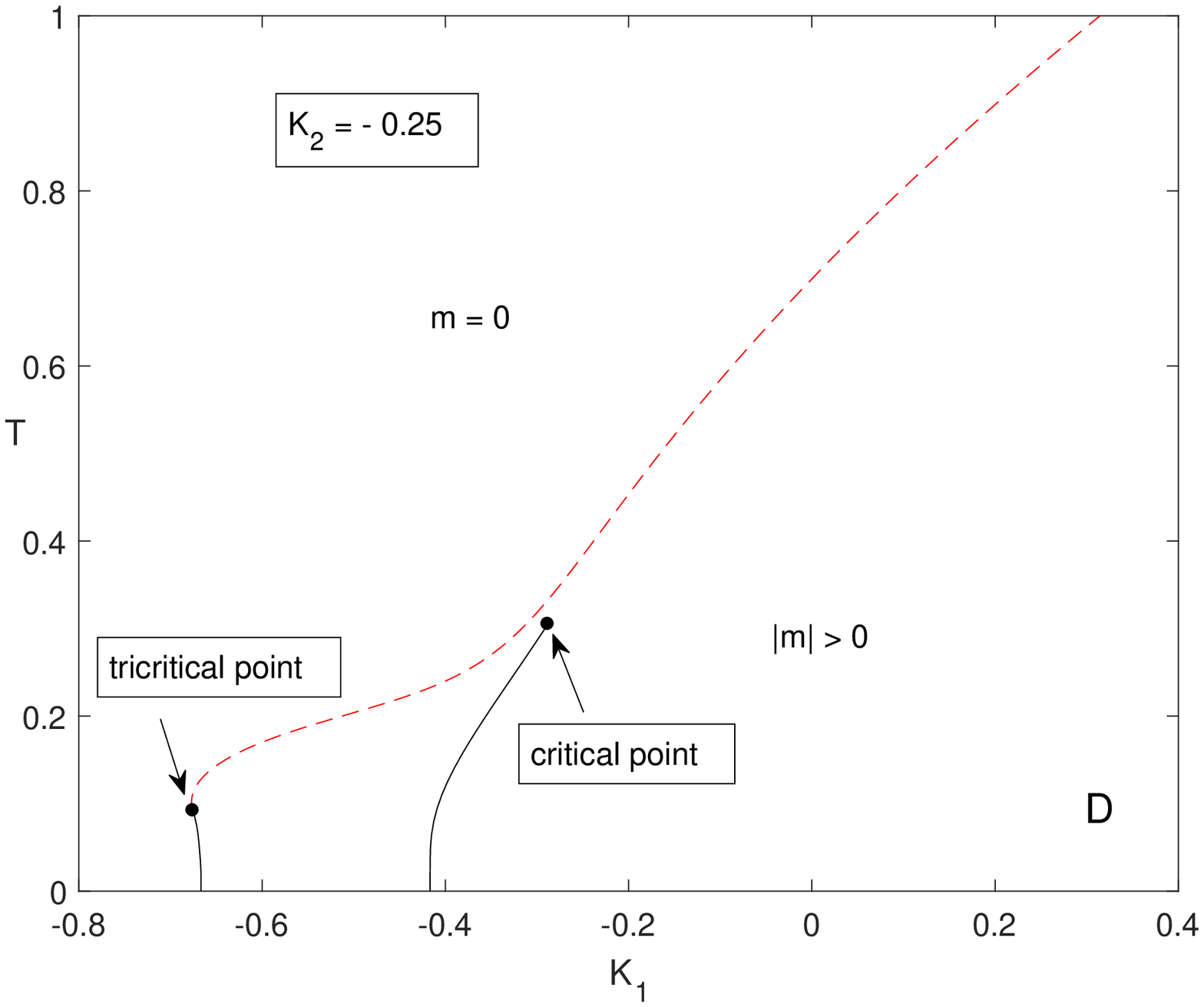}
\end{tabular}
\caption {(left)$(K_1, T)$ phase diagram for $K_2=-0.16$ in the region $C$.
(right) Phase diagram for $K_2=-0.25$ in the region $D$.} \label{Phase2}
\end{center}
\end{figure}

A typical phase diagram for a value
of $K_2$ in the region D ($K_2=-0.25$) is reported in Figure
\ref{Phase2} (right); the coordinates of the
tricritical point and of the critical point are
($K_{1} \backsimeq -0.676, T  \backsimeq 0.092$) and
($K_{1} \backsimeq -0.288, T  \backsimeq 0.305$), respectively.

The behaviour of the line $''b''$ can be defined {\it re--entrant}, in the sense
that there is a region of values of $K_1$ in which fixing $K_1$ and increasing
$T$ starting from $T=0$ the system exhibits two phase transitions,
in this case one of the first--order from vanishing $m$ to positive $|m|$ and
the other of second--order. Re--entrant phase diagrams were discussed and found
in Josephson junction arrays \cite{Fazio00}. The presence of re--entrance
in both models, despite the obvious differences between the quantum phase model
describing Josephson networks and the classical
$J$-$K_1$-$K_2$ model studied here,
traces back its common element in the
presence of NNN terms in the respective Hamiltonians.
Indeed, if the interaction term is diagonal
in the quantum phase model, there is apparently no re--entrance,
as one can see by mean--field and self--consistent harmonic
approximations \cite{Fazio00,Grignani00,Simanek94,Smerzi04}.
At variance, if there is a non--diagonal capacitance matrix, giving rise to nonlocal terms,
including a NNN coupling, then re--entrance occurs, as confirmed by Monte Carlo
simulations \cite{Capriotti03}.

We will see that the other regions E-H similarly display re--entrant behaviours.

\subsubsection{Region E:
$ -0.2745 \simeq K_2^d <K_2 < K_2^c \simeq -0.2672$\\}\label{regE}

When the critical value $K_2^c \simeq -0.2672$ is reached, further
increasing $K_2$ in modulus the
second--order line seen in region $D$ breaks in two pieces: in fact, a first--order line, limited by two tricritical points, appears;
this line is denoted by $''c''$ in the following. So in total
one has three first--order lines; the other two, $''a''$ and $''b''$, the leftmost
of which again showing re--entrant
behaviour, are qualitatively as in Region D.

\begin{figure}[!h]
\begin{center}
\begin{tabular}{ccc}
\includegraphics[width=8cm]{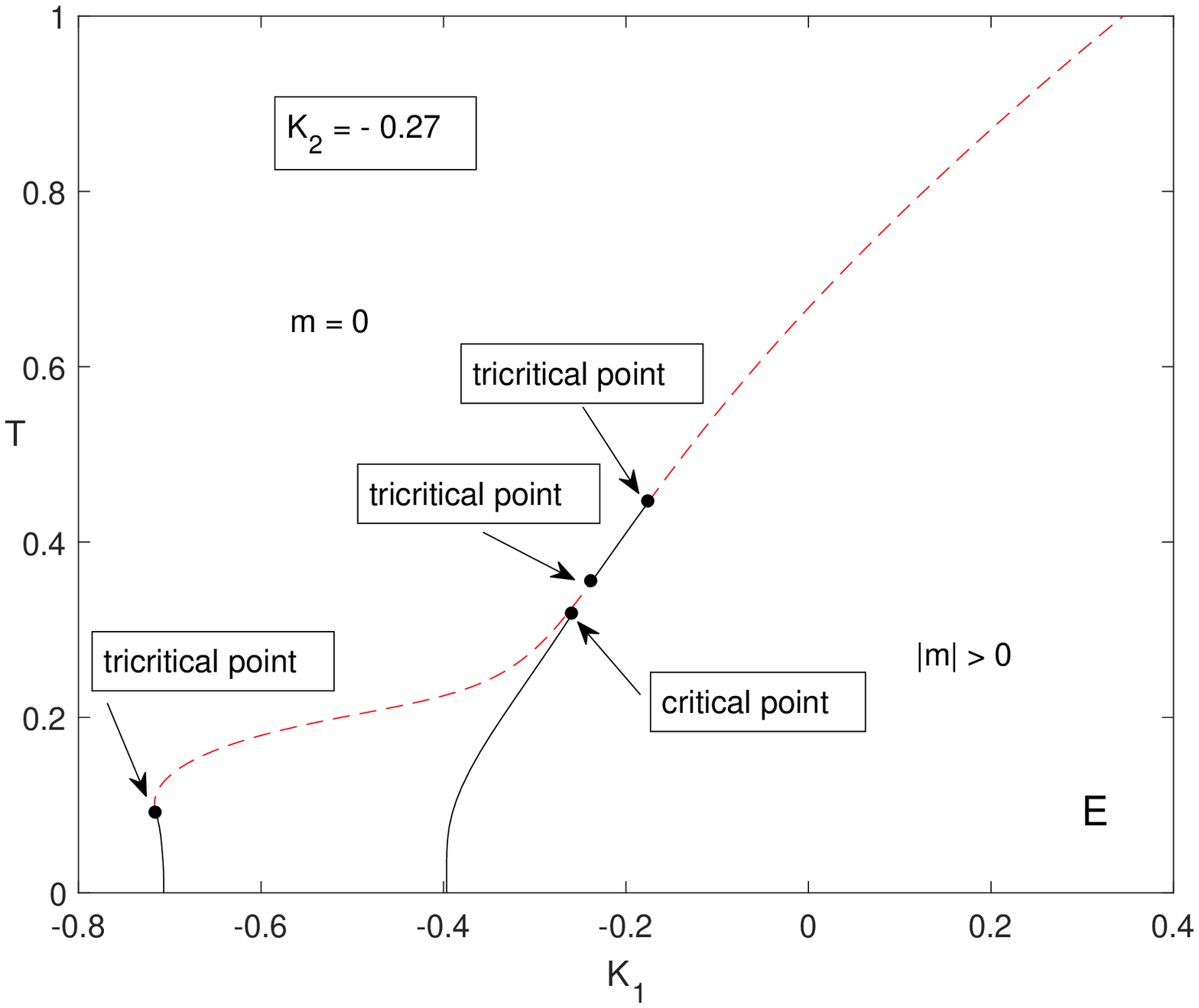} &
\includegraphics[width=8cm]{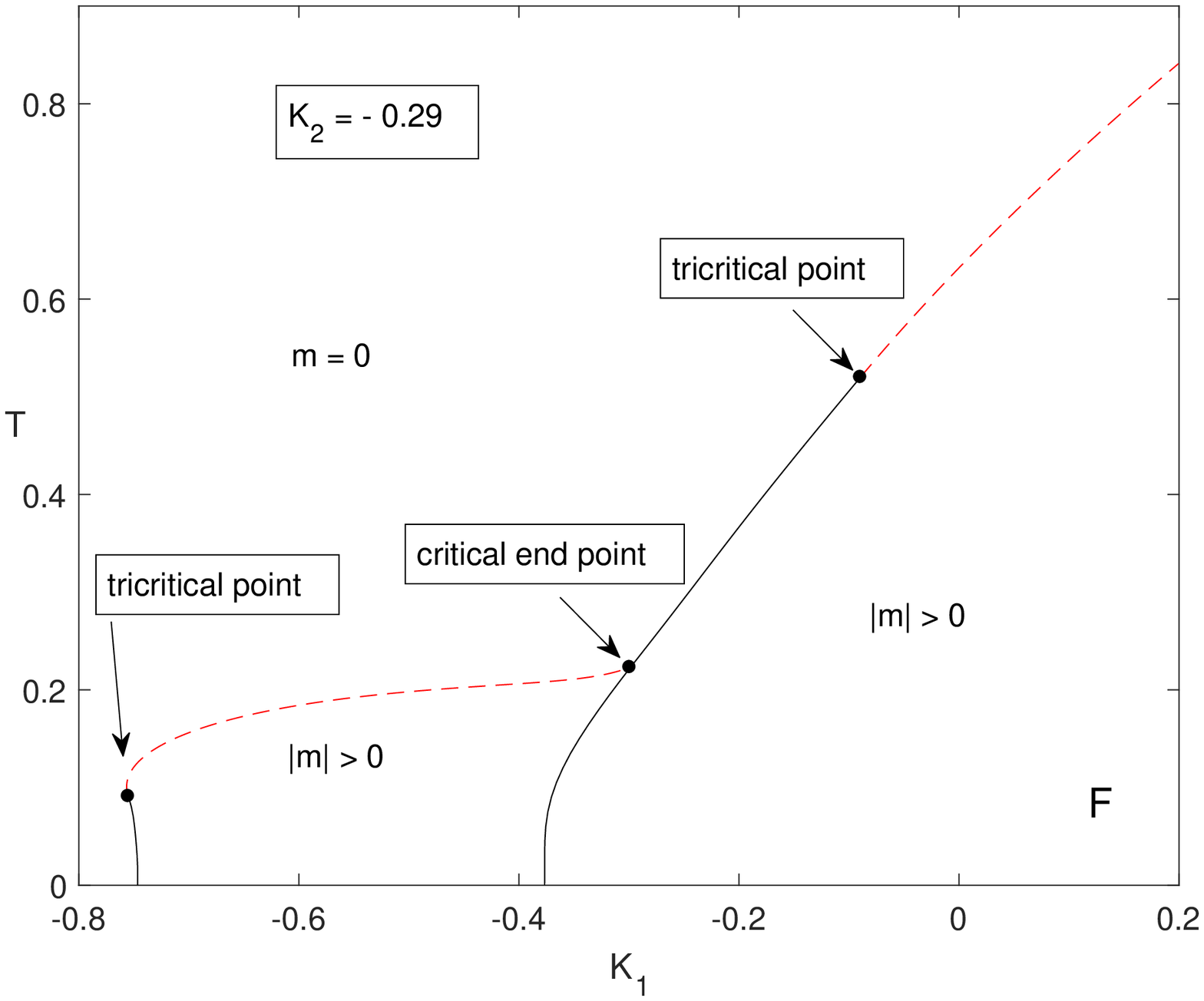}
\end{tabular}
\caption {Phase diagram for $K_2=-0.27$ in the region $E$ (left)
and for $K_2=-0.29$ in the region $F$ (right). In the
left figure the critical point is very close, but not
on, the second--order order line connecting two tricritical points.} \label{Phase3}
\end{center}
\end{figure}

The phase diagram for the value $K_2=-0.27$ in the
region E is plotted in Figure \ref{Phase3} (left), where
the coordinates of the three tricritical points are
$(K_{1} \backsimeq -0.715, T  \backsimeq 0.091)$,
($K_{1} \backsimeq -0.238, T  \backsimeq 0.355$) and
($K_{1} \backsimeq -0.175, T  \backsimeq 0.446$), while
the critical point is at $(K_{1} \backsimeq -0.259, T  \backsimeq 0.318)$.
Note that in Figure \ref{Phase3} (left) the critical point is very close, but not
on, the second--order order line connecting the two tricritical points having
the smaller temperatures.

\subsubsection{Region F:
$ -0.2953 \simeq K_2^e <K_2 < K_2^d \simeq -0.2745$\\}\label{regF}

In region E, increasing the $K_2$ in modulus, the length
of the first--order line $''c''$ (as denoted in \ref{regE})
between the two tricritical points with higher temperatures, and connected
to the two second--order lines, increases.
When the value $K_2^d \simeq -0.2745$ is reached, this first--order line
$''c''$ merges with the first--order line $''a''$ introduced in region D and present
as well in region E. Therefore a second--order line
connects the re-entrant line $''b''$ to this merged first--order line, with the
latter ending in a tricritical point and featuring a critical end point.
In Figure \ref{Phase3} (right), the phase diagram for the value
$K_2=-0.29$ in the region F is reported. The coordinates
of the two tricritical points
are $(K_{1} \backsimeq -0.755, T \backsimeq 0.091)$ and
($K_{1} \backsimeq -0.090, T  \backsimeq 0.520$), while the
critical end point is $(K_{1} \backsimeq -0.299, T  \backsimeq 0.223)$.

\subsubsection{Region G: $ -1/3 <K_2 < K_2^e \simeq -0.2953$\\}
\label{regG}

Passing from region B to region C, a first--order line bifurcates and a triple
point emerges. The same happens when the value
$K_2^e \simeq -0.2953$ is reached, entering the region G. In such a region
the line $''a''$ bifurcates in a triple point, and two first--order lines
starts from there, ending at two tricritical points that separates them
from two second--order lines. Interestingly, below the triple point
the first--order line separates two regions with non vanishing
magnetization.

The phase diagram for the value $K_2=-0.31$ in the
region G is plotted in Figure \ref{Phase4} (left), where
the triple point is $(K_{1} \backsimeq -0.325, T  \backsimeq 0.146)$,
while the three tricritical points are at
$(K_{1} \backsimeq -0.795, T  \backsimeq 0.091)$,
$(K_{1} \backsimeq -0.354, T  \backsimeq 0.167)$ and
$(K_{1} \backsimeq -0.023, T  \backsimeq 0.564)$.

\subsubsection{Region H: $  K_2 < -1/3$\\}\label{regH}

\begin{figure}[!h]
\begin{center}
\begin{tabular}{ccc}
\includegraphics[width=8cm]{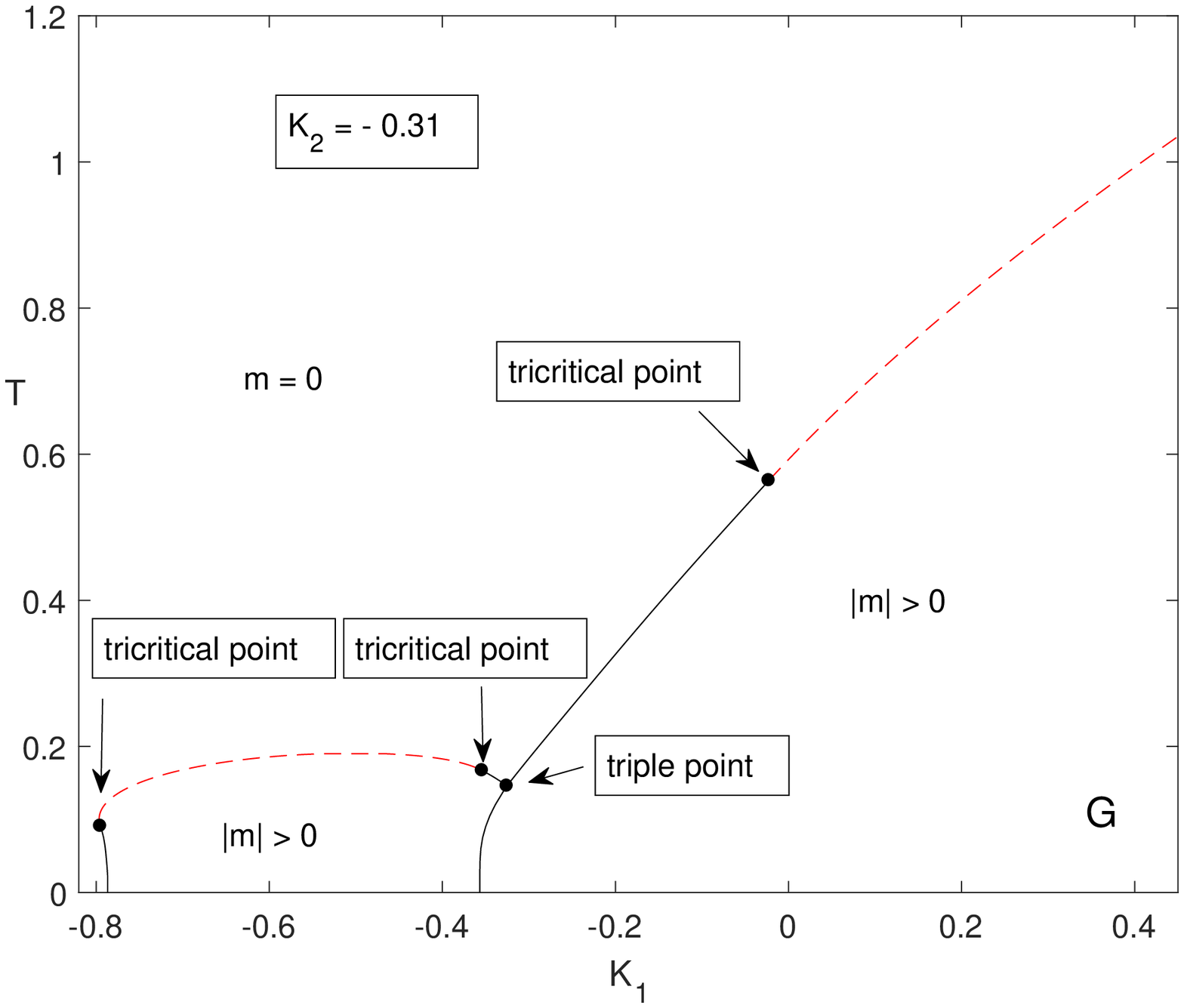} &
\includegraphics[width=8cm]{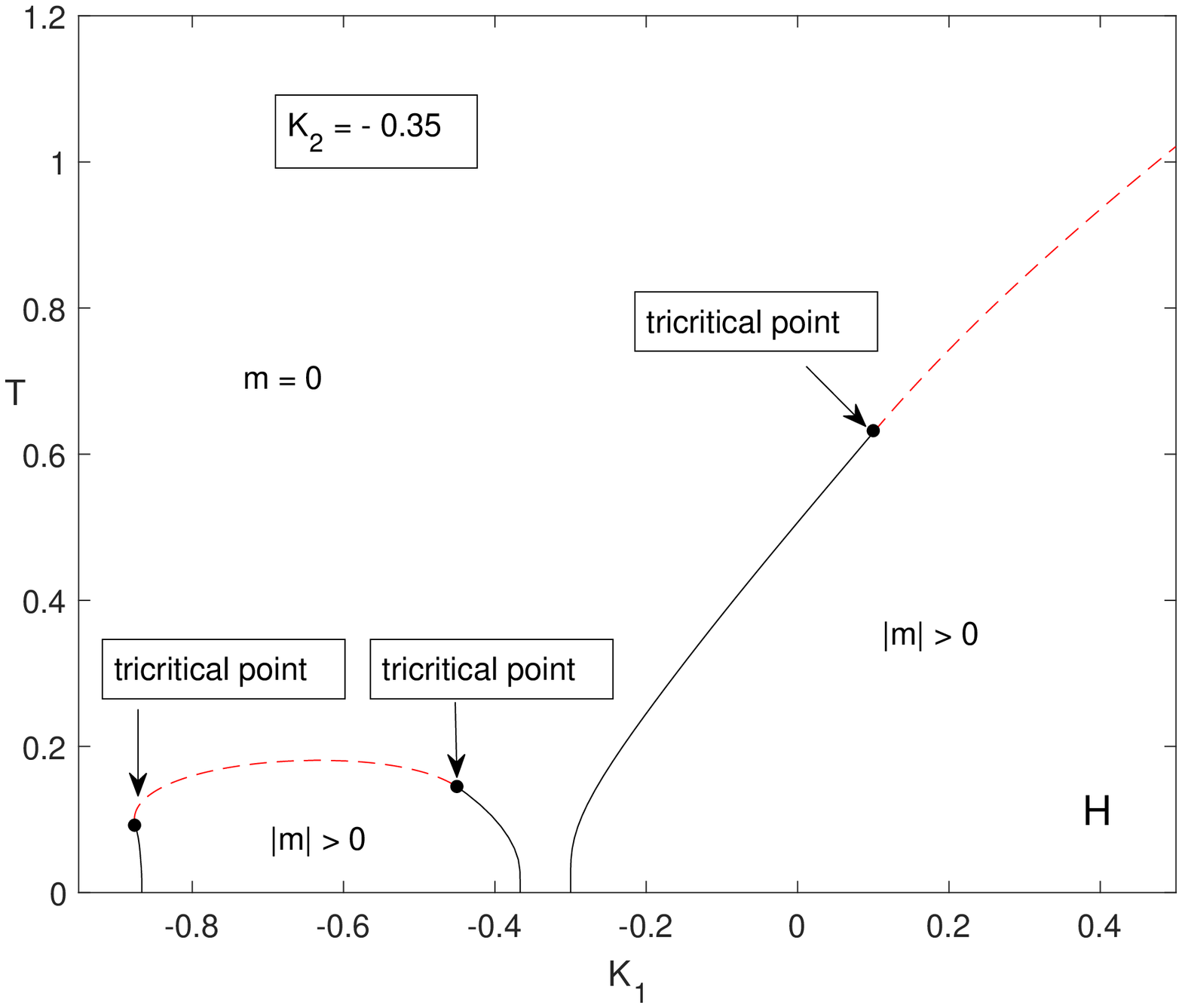}
\end{tabular}
\caption {Phase diagram for $K_2=-0.31$ in the region $G$
(left) and for $K_2=-0.35$ in the region $H$.} \label{Phase4}
\end{center}
\end{figure}

When the value of $K_2$ reaches $K_2 = -1/3$, the first--order line $''a''$ does not any longer
separate two regions with non vanishing magnetization, but actually the region
with zero magnetization persists at very low temperature arriving
to $T=0$ (region H). This part of the phase diagram is surrounded by two first--order
lines ending in tricritical points, from which two second--order lines
depart. The leftmost of them reaches at a tricritical point the third, pre--existing, first--order
line. As one can see in Figure \ref{Phase4} (right), therefore a ``lobe''
with non vanishing magnetization forms in the zero magnetization. This
lobe features two first--order lines ending in two tricritical points,
separated by a second--order line. Moreover, the line with a tricritical
point separating a first--order and a second--order line, as seen in
region A, is present for larger values of $K_1$.

Figure \ref{Phase4} (right) shows the phase diagram for the value
of $K_2=-0.35$ in the region H, where the three tricritical points are
at ($ K_{1} \backsimeq -0.875, T  \backsimeq 0.091)$,
$(K_{1} \backsimeq -0.449, T  \backsimeq 0.144)$ and
$(K_{1} \backsimeq 0.101, T  \backsimeq 0.631$).

Further decreasing $K_2$ one has that the lobe moves towards left, but apart
from that no qualitative changes in the phase diagram are expected. Notice
that when both $K_1$ and $K_2$ becomes large in modulus,
this corresponds to have
$J$ going to zero (since we are putting $J=1$),
and in this limit the transition
occurs only at lower and lower temperatures. A similar situation occurs for
$K_1$ and
$K_2$ positive and large, so that we expect that in region A no qualitative
changes occurs increasing, {\it e.g.}, $K_2$.

Finally, we observe that if one fixes $K_1$ and look at the
phase diagrams in the $(K_2,T)$ plane, one expects to find again
the same singularities; we defer a discussion on this point to the next
Section.

\section{Discussion}
\label{discussion}

The classification of phase transitions
in systems with long--range interactions \cite{Bouchet05} shows that the
structure of phase diagrams can be quite complex. The model
studied in this work provides a concrete example of this fact.
The Hamiltonian has two parameters,
$K_1$ and $K_2$, and therefore the phase diagram lies in the three--dimensional
space
$(K_1,K_2,T)$. We have chosen to plot two--dimensional cuts of the phase diagram, defined
by different fixed values of $K_2$. This was due to the usefulness to have a simple
visualization of the diagram, but at the same time this provided a comparison
with the rather simple two--dimensional phase diagram of the model with only the NN
coupling $K_1$ (i.e., with $K_2=0$). The complexity
of the phase diagram has resulted in a
quite rich behaviour of these two--dimensional plots. In fact,
eight qualitatively different structures of two--dimensional phase diagrams are present, each one with
its own structure of first-- and second--order phase transition lines and
points that in the plots occur in four different types: tricritical points,
triple points, critical end points, and critical points. For brevity, in this discussion we refer
collectively to these four types of points as relevant points.

Each one of the eight different structures occurs for a given range of $K_2$, and in the previous Section
we have shown a diagram for each structure.
In the following we provide a qualitative description of the changes occurring when moving from one structure
to another in terms of the change in the occurrence of relevant points.
Afterwards, we will give a more general view based on the classification of singularities that can be
present in the phase diagrams \cite{Bouchet05}.
The passages are indicated in the following list.

\begin{itemize}

\item A tricritical point splitting in a pair consisting of a critical end point
and a critical point. This is seen passing through the value of $K_2$
given by $K_2=K_2^a \simeq -0.0885$, {\it i.e.} passing from region A to region B, see the two panels
of Figure \ref{Phase1}.

\item A critical end point splitting in a pair consisting of a triple point and a
tricritical point.  This is seen at the value $K_2=K_2^b \simeq -0.1542$,
passing from region B to region C, as seen in Figures \ref{Phase1}
(right) and \ref{Phase2} (left), and at the value $K_2=K_2^e \simeq -0.2953$,
passing from region F to region G, see Figures \ref{Phase3}
(right) and \ref{Phase4} (left).

\item A triple point reaching $T=0$
and giving rise to a pair of first--order lines starting at $T=0$, and
separated (at $T=0$) by a region in which the magnetization is not vanishing.
This is seen from region C to region D, at $K_2=-1/6$, see the two panels of Figure
\ref{Phase2}. Also at $K_2=-1/3$, as illustrated in the comparison of the two panels of
Figure \ref{Phase4}, showing the passage from region G to region H, there is an increase,
in this case from $2$ to $3$, of the number of
first--order lines starting from $T=0$.

\item A first--order line, between two emerging tricritical
points, generated in a second--order line. This is seen in
Figures \ref{Phase2} (right) and \ref{Phase3} (left), with the passage
from region D to region E occurring at $K_2=K_2^c \simeq - 0.2672$.

\item A pair consisting of a tricritical point and a critical point merging
in a critical end point. This is seen passing from region E to region F, at
$K_2=K_2^d \simeq -0.2745$, as one can notice in the two panels of Figure
\ref{Phase3}.

\end{itemize}

\subsection{Classification of singularities}
\label{class_sing}

If the Hamiltonian of a model has $p$ parameters (in our case $p=2$), its thermodynamic phase
diagram has $p+1$ dimensions, corresponding to the parameters plus the temperature $T$. In the
phase diagram there can be singularities that span hypersurfaces of dimension $0$ (points),
$1$ (lines), $2$ (surfaces), and so on up to $p$; therefore in our case we can have singularities
spanning points, lines and surfaces in a three--dimensional space. A singularity is said to be
of codimension $n$ if it spans a hypersurface of dimension $p-n$; thus here we have singularities
of codimension $0$ (surfaces), $1$ (lines) and $2$ (points). In Ref. \cite{Bouchet05} a complete
classification of codimension--$0$ and codimension--$1$ singularities, in systems with long--range
interactions, has been given. Our results can be discussed in that framework.

Clearly the first-- and second--order phase transition lines of our
two--dimensional plots correspond
to surfaces in the three--dimensional $(K_1,K_2,T)$ space and are codimension--$0$ singularities, while
the relevant points in our plots correspond to lines in the $(K_1,K_2,T)$ space and are codimension--$1$ singularities.
Furthermore, the codimension--$1$ lines result from the intersection of two codimension--$0$ surfaces. It is then not
difficult to see that codimension--$2$ singularities are the points
resulting from the intersection of three surfaces,
and in fact in a three--dimensional space
three surfaces {\it generically} (i.e., apart from particular cases)
meet in a point. Such a point can also be seen as the point where the
three lines defined by the intersection of
the three couples of surfaces, formed out of the three surfaces, converge.
It must be stressed that each one of the
three lines converge to the point only from one direction,
since beyond the point each line would correspond to non--equilibrium
(unstable or metastable) states.

In this respect, the meeting and splitting of relevant points described
in the list above, is the description
of the convergence of codimension--$1$ singularities (lines) in a
codimension--$2$ singularity (point) (with
an exception that we treat below). In order for a
codimension--$2$ singularity to appear in one of our two--dimensional
plots at fixed $K_2$, we should have made the plot for
exactly the $K_2$ value where the singularity occurs, and the
$K_2$ values of these singularities are the range boundaries that we have
specified in the previous section
and in the list above. On the basis of these arguments, one can argue the
following. Had we chosen another way
to present our results in two--dimensional plots, {\it e.g.},
by plotting $(K_2,T)$ diagrams at various fixed values of $K_1$,
or by choosing other more complicated two--dimensional cuts defined by various fixed values of the quantity
$a K_1 + b K_2$ with given $a$ and $b$,
we still would have observed a qualitative change of structure at the passage
of the plane through the $(K_1,K_2,T)$ points where the
codimension--$2$ singularities are located. However,
this qualitative changes would have been characterized, in general, by meetings and splittings of relevant
points different from those given in the list above.
The conclusion is that apart from the last details,
we would have observed the same sequence of qualitative change of structures,
and the same richness of such structures,
for any choice of two--dimensional cuts of the phase diagram.
We observe that the appearance of two
tricritical points (the passage from region D to region E) is not
due to the crossing of a codimension--$2$ singularities, but
it is the result of the following fact.
In region E the plane with constant $K_2$ crosses the line of
tricritical points
in two distinct points of the $(K_1,T)$ plane; approaching the
boundary with region D the two points approach each other, until
at $K_2=K_2^c$ the plane is tangent to the line of tricritical points and the two points in the $(K_1,T)$ plane coincide. When
$K_2$ enters the range of region D there is no more an intersection,
and there are no more tricritical points in the $(K_1,T)$ plane.
Although this qualitative change is not related to a codimension--$2$
singularity, it still would manifest itself in another choice
of two--dimensional cuts, since the same mechanism would occur.

It is clear that adding couplings acting at
larger distances, such $K_3$, $K_4$ and so on, the phase diagram would
have further
dimensions and would become very complex. For example, with just
the presence of $K_3$ the phase diagram would be four--dimensional, and
one could, {\it e.g.}, represent it with three--dimensional cuts
at various fixed values of $K_3$. From
the results presented in this work one can guess that the number of qualitatively different structures of such three--dimensional plots,
varying $K_3$, would be quite large, and the passage from one structure to the others, due to the crossing of codimension--$3$ singularities,
would be characterized by an extremely rich set of possibilities.
This shows that this class of systems with finite--range coupling
in presence of a long--range term appears to be an ideal playground
to see in simple, yet meaningful, models, the behaviour of thermodynamical
singularities and the relation between different types of
critical points.

\section{Conclusions}
\label{conclusions}

In this paper we studied an Ising
spin chain with short--range competing interactions in presence of
long--range couplings. We worked out
the partition function of the model and the phase diagram in the canonical
ensemble. We found that eight
possible, qualitatively different two--dimensional phase diagrams exist
in the parameter space. They occur as a result
of the frustration and the competition
between the short-- and the long--range interactions.

One of the motivations of our work was that, when one of the two short--range
interactions is turned off, the system exhibits ensemble inequivalence
and the thermodynamic and dynamical behavior of the system
in both the canonical and microcanonical ensembles may be different. Our
study is a first step in the investigation of the effects of additional
finite--range coupling terms, since we provided a full characterization
of the canonical phase diagrams. Therefore,
it is appealing to consider the solution of the model studied
in this paper in the microcanonical ensemble. For such a study one could
apply the method of determining the entropy presented in
Ref. \cite{Gori11}. One can
anticipate that a very rich microcanonical behaviour occurs when $K_2$
is added. This can be argued from the fact that with $K_2=0$ the canonical phase
diagram has only a tricritical point, while for $K_2 \neq 0$ one has all the
possibilities described in Section \ref{phasediag}.
It would be then very interesting to work out the details of the
comparison between the two ensembles, and study their inequivalence
when $K_2$ is turned on.

Several extensions of the model studied here could be as well very interesting. One could study, in the same one--dimensional geometry considered in the
present paper: {\em i)} the effect of short--range interactions in presence
of mean-field terms for $O(n)$ models
\cite{Campa03,Campa06,Dauxois10},
{\em ii)} spin--$1$ systems \cite{Hovhannisyan17},
{\em iii)} more general long--range interactions
with power--law decay \cite{Campa14}, and {\em iv)} the effect
of a short--range term in the Sherrington-Kirkpatrick spin glass model
\cite{Mezard87}. It would be also appealing to consider our model
in higher dimensions, since it is known that already in two dimensions one can
have antiferromagnetic phases at finite temperature \cite{Kardar_PRL}.
One could also ask whether it is possible to enlarge the re--entrance
in the phase diagram by having short--range terms involving more
couplings. Finally, it would be very deserving to study the quantum version
of the classical
model considered here and compare the results with other
quantum models with competing short-- and long--range interactions
\cite{Igloi18}.

\ack
The authors thanks N. Defenu and D. Mukamel for very useful discussions.
A.C. acknowledges financial support from INFN (Istituto Nazionale di Fisica Nucleare) through the projects DYNSYSMATH and ENESMA.
V.H. acknowledges financial support from the RA MES
State Committee of Science, in the frames of the research project No. SCS 18T-1C155.

This paper honors the $70$th birthday
of Giorgio Parisi; S.R. thanks Giorgio
for introducing him to the amazing field of statistical mechanics and for being
always a source of inspiration from both the scientific and the human side.

\appendix
\section{The evaluation of the ground state}
\label{eval_app}

For convenience we reproduce here the expressions of the energy per
particle and the allowed ranges of the
order parameters:
\be
\label{enerorder_app}
\epsilon = - \frac{1}{2} \left( m^2 + K_1 g_1 + K_2 g_2 \right) \, ,
\ee
\be
\label{order_range_app}
\mkern-36mu \mkern-36mu
-1 \leq m \leq 1, \,\,\,\,\,\,\,\,\, 2|m|-1 \leq g_1 \leq 1, \,\,\,\,\,\,\,\,\,
\max(2|m|-1,2|g_1|-1) \leq g_2 \leq 1 \, .
\ee
The evaluation of the order parameter values giving the ground state can be divided in four steps, corresponding respectively
to the four quadrants of the $(K_1,K_2)$ plane.
\subsection*{{\rm I)} $K_1\geq 0$, $K_2 \geq 0$}
The state of lowest energy is obtained when all three order parameters are equal to $1$,
i.e., with a fully magnetized system.
\subsection*{{\rm II)} $K_1 < 0$, $K_2 \geq 0$}
For given $m$ and $g_1$ the lowest energy is achieved for $g_2=1$, giving
\be
\label{ener_case2}
\epsilon = -\frac{1}{2} \left( m^2 + K_1 g_1 + K_2 \right) \, .
\ee
Since $K_1$ is negative, for given $m$ the lowest value of this expression is obtained for the smallest allowed value of $g_1$.
This value, restricting to $m\geq 0$ (as we have explained can be done), is $2m-1$. Then we have to find the smallest value
of
\be
\label{ener_case2b}
\epsilon = -\frac{1}{2} \left( m^2 +2K_1 m -K_1 +K_2 \right)
\ee
for $0\leq m \leq 1$. One finds right away that this is obtained for $m=1$ (and thus $g_1=1$) for $K_1 > -\frac{1}{2}$, and for
$m=0$ (and thus $g_1=-1$) for $K_1 < - \frac{1}{2}$.
\subsection*{{\rm III)} $K_1 \geq 0$, $K_2 < 0$}
The two cases with $K_2<0$ are less immediate. When $K_1 \geq 0$ we can proceed as follows. From the inequalities (\ref{order_range_app})
one deduces that $g_1 \leq (1+g_2)/2$. Then, since $K_1 \geq 0$, for given $m$ and $g_2$ the lowest value of the energy per particle
(\ref{enerorder_app}) is obtained for $g_1=(1+g_2)/2$, giving
\be
\label{ener_case3}
\epsilon = -\frac{1}{2} \left[ m^2 + \left( \frac{1}{2}K_1 +K_2\right) g_2 +\frac{1}{2} K_1 \right] \, .
\ee
If $\frac{1}{2}K_1 + K_2 > 0$ the minimum of this expression is obtained for $g_2=1$ (and thus $g_1=1$) and $m=1$. If
$\frac{1}{2}K_1 + K_2 < 0$ the minimum for given $m$ is achieved for $g_2=2m-1$ (and thus $g_1=m$), to have
\be
\label{ener_case3b}
\epsilon = -\frac{1}{2} \left[ m^2 + \left( K_1 + 2K_2\right) m - K_2 \right] \, .
\ee
The minimum of this expression (for $0\leq m \leq 1$) occurs for $m=0$ (and thus $g_2=-1$ and $g_1=0$) when $K_2 < -\frac{1}{2} K_1 -\frac{1}{2}$,
and for $m=1$ (and thus $g_2=1$ and $g_1=1$) when $K_2 < -\frac{1}{2} K_1 -\frac{1}{2}$. Considering also the situation in which
$\frac{1}{2}K_1 + K_2 > 0$, the last two expression give the overall result for this case ($K_1 \geq 0$ and $K_2 <0$).
\subsection*{{\rm IV)} $K_1 <0$, $K_2 <0$}
This is the case requiring more attention.
It is convenient to divide the range $0 \leq m \leq 1$ in the two subranges
$0 \leq m \leq 1/3$ and $1/3 \leq m \leq 1$. Let us begin with the latter.

$a$) $1/3 \leq m \leq 1$. The smallest allowed value of $g_1$, i.e., $2m-1$, can be negative if $m< 1/2$. However, regardless of this possibility,
for $m \geq 1/3$ it will always be $-(2m-1) \leq m$; therefore, for $2m-1 \leq g_1 \leq m$ it will always be $m\geq |g_1|$. Consequently, for given
$m$ in this subrange and given $g_1$ between $2m -1$ and $m$, the lowest value of the energy (\ref{enerorder_app}) is obtained for $g_2=2m-1$.
On the other hand, when $m\leq g_1 \leq 1$ the lowest value occurs for $g_2=2g_1 -1$. After making these positions and seeking for the lowest value
varying $g_1$, this will occur for $g_1=2m-1$, and thus also $g_2=2m-1$. We then obtain the expression
\be
\label{ener_case4}
\epsilon = -\frac{1}{2} \left[ m^2 + 2 \left( K_1 +K_2\right) m - K_1 - K_2 \right] \, .
\ee
The minimum of this expression in the range $1/3 \leq m \leq 1$ is obtained for $m=1$ (and thus $g_1=1$ and $g_2=1$) when $K_2 > -K_1 -\frac{2}{3}$,
and for $m=1/3$ (and thus for $g_1 = -1/3$ and $g_2 = -1/3$) when $K_2 < -K_1 -\frac{2}{3}$ (reminding that these bounds have to be considered together
with $K_2 <0$). The corresponding minima will have to be compared with the minima in the range $0 \leq m \leq 1/3$, to find the overall minima in $m$.

$b$) $0 \leq m \leq 1/3$. Now $-(2m-1)$ is larger than $m$. Therefore, we can repeat the above analysis only for $-m \leq g_1 \leq 1$. As a consequence,
the lowest value of the energy for given $m$ in this subrange and for given $g_1$ larger than $-m$ is obtained for $g_1=-m$ and $g_2=2m-1$, giving the expression
\be
\label{ener_case4b}
\epsilon = -\frac{1}{2} \left[ m^2 - \left( K_1 - 2 K_2\right) m - K_2 \right] \, .
\ee
For $2m-1 \leq g_1 \leq -m$, $|g_1|$ is larger than $m$, then the minimum occurs for $g_2= 2|g_1| -1 = -2g_1 -1$. Then
\be
\label{ener_case4c}
\epsilon = -\frac{1}{2} \left[ m^2 + \left( K_1 - 2 K_2\right) g_1 - K_2 \right] \, .
\ee
For $K_1 -2K_2 >0$ the minimum in $g_1$ occurs for $g_1=-m$ (and thus $g_2=2m-1$), while for $K_1 -2K_2 <0$ the minimum in $g_1$
occurs for $g_1=2m-1$ (and thus $g_2=-4m+1$). We then obtain the expressions:
\bea
\label{ener_case4d}
\epsilon &=& -\frac{1}{2} \left[ m^2 - \left( K_1 -2K_2 \right) m -K_2 \right]; \,\,\,\,\,\,\,\,\,\,\,\,\,\,\, K_1 -2K_2 >0 \\
\label{ener_case4e}
\epsilon &=& -\frac{1}{2} \left[ m^2 +2\left( K_1 -2K_2 \right) m -K_1 +K_2 \right]; \,\,\,\,\,\,\,\,\,\,\,\,\,\,\, K_1 -2K_2 <0
\eea
Studying these expressions in the range $0 \leq m \leq 1/3$ one finds the minimum occurs for $m=0$, $g_1=0$ and $g_2=-1$ for
$K_2 < \frac{1}{2}K_1 - \frac{1}{6}$; it is found for $m= 1/3$, $g_1=-1/3$ and $g_2=-1/3$ for $\frac{1}{2}K_1-\frac{1}{6}
<K_2 < \frac{1}{2}K_1 + \frac{1}{12}$; it occurs for $m=0$, $g_1=-1$ and $g_2=1$ for $K_2 > \frac{1}{2}K_1 + \frac{1}{12}$ (again, all these bounds have
to be considered together with $K_2 <0$).

Comparing the subcases $a$) and $b$) we obtain the configurations with lowest energy in the case IV, i.e., in the quadrant $K_1 <0$, $K_2<0$. We find
the following results. For $-\frac{1}{3} \leq K_1 \leq 0$ the minimum is at $m=1$, $g_1=1$ and $g_2=1$ for $K_2 > -\frac{1}{2}K_1 - \frac{1}{2}$, and at
$m=0$, $g_1=0$ and $g_2=-1$ for $K_2 < -\frac{1}{2}K_1 -\frac{1}{2}$. For $-\frac{1}{2} \leq K_1 \leq -\frac{1}{3}$ the minimum is at
$m=0$, $g_1=0$ and $g_2=-1$ for $K_2 < \frac{1}{2}K_1 -\frac{1}{6}$, at $m=1/3$, $g_1=-1/3$ and $g_2=-1/3$ for $\frac{1}{2}K_1 - \frac{1}{6}
<K_2 < -K_1 -\frac{2}{3}$, and at $m=1$, $g=1$ and $g_2=1$ for $K_2 > -K_1 -\frac{2}{3}$. For $K_1 \leq -\frac{1}{2}$ the minimum is at
$m=0$, $g_1=0$ and $g_2=-1$ for $K_2 < \frac{1}{2}K_1 -\frac{1}{6}$, at $m=1/3$, $g_1=-1/3$ and $g_2=-1/3$ for $\frac{1}{2}K_1 - \frac{1}{6}
<K_2 < \frac{1}{2}K_1 +\frac{1}{12}$, and at $m=0$, $g_1=-1$ and $g_2=1$ for $K_2 > \frac{1}{2}K_1 + \frac{1}{12}$.

At the end, unifying all the cases from I to IV, one obtains, for the ground state, the results described in Section \ref{groundstate}
and shown in Figure \ref{fig_ground}.


\section*{References}
\begin{thebibliography}{99}

\bibitem{Mezard87}
M. Mezard, G. Parisi, and M. A. Virasoro, {\it
Spin glass theory and beyond} (Singapore, World Scientific, 1987).

\bibitem{Chaikin95}
P. M. Chaikin and T. C. Lubensky,
{\it Principles of condensed matter physics}
(Cambridge, Cambridge University Press, 1995).

\bibitem{Seul95}
M. Seul and D. Andelman, Science {\bf 267}, 476 (1995).

\bibitem{Giuliani09}
A. Giuliani, J. L. Lebowitz, and E. H. Lieb, AIP Conference Proceedings
{\bf 1091}, 44 (2009).

\bibitem{Chakrabarty11}
S. Chakrabarty and Z. Nussinov, Phys. Rev. B {\bf 84}, 144402 (2011).

\bibitem{Diep04}
H. T. Diep, {\it Frustrated spin systems} (Singapore, World Scientific, 2004).

\bibitem{Redner81}
S. Redner, J. Stat. Phys. {\bf 25}, 15 (1981).

\bibitem{Mila91}
F. Mila, D. Poilblanc, and C. Bruder,
Phys. Rev. B 43, 7891 (1991).

\bibitem{Singh99}
R. R. P. Singh, W. H. Zheng, C. J. Hamer, and J. Oitmaa,
Phys. Rev B {\bf 60}, 7278 (1999).

\bibitem{Capriotti04}
L. Capriotti, F. Becca, A. Parola, and S. Sorella,
Phys. Rev. Lett. {\bf 87}, 097201 (2001).

\bibitem{Sirker06}
J. Sirker, Z. Weihong, O. P. Sushkov, and J. Oitmaa,
Phys. Rev. B {\bf 73}, 184420 (2006).

\bibitem{Spenke12}
M. Spenke and S. Guertler,
Phys. Rev. B {\bf 86}, 054440 (2012).

\bibitem{Wang18}
L. Wang and A. W. Sandvik, Phys. Rev. Lett. {\bf 121}, 107202 (2018).

\bibitem{Dyson69}
F. J. Dyson, Comm. Math. Phys. {\bf 12}, 91 (1969).

\bibitem{Thouless69}
D. J. Thouless, Phys. Rev. {\bf 187}, 732 (1969).

\bibitem{Sak73}
J. Sak, Phys. Rev. B {\bf 8}, 281 (1973).

\bibitem{Luijten02}
E. Luijten and H. W. J. Bl\"ote,
Phys. Rev. Lett. {\bf 89}, 025703 (2002).

\bibitem{Blanchard13}
T. Blanchard, M. Picco, and M. A. Rajabpour,
Europhys. Lett. {\bf 101}, 56003 (2013).

\bibitem{Brezin14}
E. Brezin, G. Parisi, and F. Ricci-Tersenghi,
J. Stat. Phys. {\bf 157}, 855 (2014).

\bibitem{Angelini14}
M. C. Angelini, G. Parisi, and F. Ricci-Tersenghi,
Phys. Rev. E {\bf 89}, 062120 (2014).

\bibitem{Defenu15} N. Defenu, A. Trombettoni, and A. Codello,
Phys. Rev. E {\bf 92}, 052113 (2015).

\bibitem{Defenu16} N. Defenu, A. Trombettoni, and S. Ruffo,
Phys. Rev. B {\bf 94}, 224411 (2016); Phys. Rev. B {\bf 96}, 104432 (2017)

\bibitem{Gori17}
G. Gori, M. Michelangeli, N. Defenu, and A. Trombettoni,
Phys. Rev. E {\bf 96}, 012108 (2017).

\bibitem{Behan17}
C. Behan, L. Rastelli, S. Rychkov, and B. Zan, Phys. Rev.
Lett. {\bf 118}, 241601 (2017);
J. Phys. A {\bf 50}, 354002 (2017).

\bibitem{Morita17}
T. Horita, H. Suwa, and S. Todo, Phys. Rev. E {\bf 95}, 012143 (2017).

\bibitem{Campa14}
A. Campa, T. Dauxois, D. Fanelli and S. Ruffo, {\it
Physics of long-range interacting systems} (Oxford, Oxford University Press, 2014).

\bibitem{Nagle70}
J. F. Nagle, Phys. Rev. A {\bf 2}, 2124 (1970).

\bibitem{Kardar83}
M. Kardar, Phys. Rev. B {\bf 28}, 244 (1983).

\bibitem{Mukamel05}
D. Mukamel, S. Ruffo, and N. Schreiber,
Phys. Rev. Lett. {\bf 95}, 240604 (2005).

\bibitem{Parisi_book}
G. Parisi, {\it Statistical field theory} (Redwood City, Addison-Wesley, 1988).

\bibitem{Perron07}
O. Perron, Mathematische Annalen {\bf 64}, 248 (1907).

\bibitem{Frobenius12}
F. G. Frobenius, Sitzungsber. K\"{o}nigl. Preuss. Akad. Wiss. 456 (1912).

\bibitem{Varga09}
R. S. Varga, {\it Matrix Iterative Analysis},
Springer Series in Computational Mathematics, Vol. 27 (Springer,
Berlin \& Heidelberg, 2009).

\bibitem{Fazio00}
R. Fazio and H. Van Der Zant,
Phys. Rep. {\bf 355}, 235 (2000).

\bibitem{Grignani00}
G. Grignani, A. Mattoni, P. Sodano, and A. Trombettoni,
Phys. Rev. B {\bf 61}, 11676 (2000).

\bibitem{Simanek94}
E. \u{S}im\'{a}nek, {\it
Inhomogeneous superconductors: granular and quantum effects}
(Oxford University Press, Oxford, 1994).

\bibitem{Smerzi04}
A. Smerzi, P. Sodano, and A. Trombettoni,
J. Phys. B {\bf 37}, S265 (2004).

\bibitem{Capriotti03}
L. Capriotti, A. Cuccoli, A. Fubini, V. Tognetti, and R. Vaia,
Phys. Rev. Lett. {\bf 91}, 247004 (2003).

\bibitem{Bouchet05}
F. Bouchet and J. Barr\'{e}, J. Stat. Phys. {\bf 118}, 1073 (2005).

\bibitem{Gori11}
G. Gori and A. Trombettoni, J. Stat. Mech. P10021 (2011).

\bibitem{Campa03}
A. Campa, A. Giansanti, and D. Moroni, J. Phys. A {\bf 36}, 6897 (2001).

\bibitem{Campa06}
A. Campa, A. Giansanti, D. Mukamel, and S. Ruffo,
Physica A {\bf 365}, 177 (2006).

\bibitem{Dauxois10}
T. Dauxois, P. de Buyl, L. Lori, and S. Ruffo,
J. Stat. Mech. P06015 (2010).

\bibitem{Hovhannisyan17}
V. V. Hovhannisyan, N. S. Ananikian, A. Campa, and S. Ruffo,
Phys. Rev. E {\bf 96}, 062103 (2017).


\bibitem{Kardar_PRL}
  M. Kardar, Phys. Rev. Lett. {\bf 51}, 523 (1983).
  

\bibitem{Igloi18}
F. Igl\'{o}i, B. Bla\ss, G. Ro\'{o}sz, and H. Rieger,
Phys. Rev. B {\bf 98}, 184415 (2018).

\end{thebibliography}
\end{document}